\documentclass[12pt]{iopart}
\usepackage{color}

\usepackage{graphicx} 
\usepackage{relsize}

\def\beqra{\begin{eqnarray}}
\def\eeqra{\end{eqnarray}}
\def\beq{\begin{equation}}
\def\eeq{\end{equation}}

\def\vb{\bar{v}}
\def\nb{\bar{n}}
\def\fb{\bar{f}}
\def\sigmab{\bar{\sigma}}

\def\bk{{\bf k}}
\def\bw{{\bf w}}
\def\vp{\varphi}

\def\vpb{\bar{\vp}}
\def\bx{{\bf{x}}}
\def\by{{\bf{y}}}
\def\bp{{\bf{p}}}
\def\bq{{\bf{q}}}
\def\bl{{\bf{l}}}

\def\agt{\stackrel{>}{\sim}}
\def\alt{\stackrel{<}{\sim}}
\begin{document}

\title{Coarse-Grained Cosmological Perturbation Theory}

\author{M. Pietroni$^1$, G. Mangano$^2$, N. Saviano$^3$, M. Viel$^4$}

\address{
$^1$ Istituto Nazionale di Fisica Nucleare - Sezione di
Padova - \\ Via Marzolo 8, I-35131 Padova, Italy\\
$^2$ Istituto Nazionale di Fisica Nucleare - Sezione di
Napoli - \\ Complesso Universitario di Monte S.Angelo, I-80126
Napoli, Italy\\
$^3$ II Institut f\"ur Theoretische Physik, Universit\"at Hamburg, Luruper Chaussee 149, D-22761, Hamburg, Germany \\$^4$INAF-Osservatorio Astronomico di Trieste, Via G.B. Tiepolo 11, I-34131 and \\  Istituto Nazionale di Fisica Nucleare - Sezione di Trieste, Via Valerio 2, I-34127 Trieste, Italy}

\ead{\\massimo.pietroni@pd.infn.it\\mangano@na.infn.it \\
ninetta.saviano@desy.de\\viel@oats.inaf.it}

\begin{abstract}
Semi-analytical methods, based on Eulerian perturbation theory, are a promising tool to follow the time evolution of cosmological perturbations at small redshifts and at mildly nonlinear scales. All these schemes are based on two approximations: the existence of a smoothing scale and the single-stream approximation, where velocity dispersion of the dark matter fluid, as well as higher moments of the particle distributions, are neglected. Despite being widely recognized, these two assumptions are, in principle, incompatible, since any finite smoothing scale gives rise to velocity dispersion and higher moments at larger scales.

We describe a new approach to perturbation theory, where the Vlasov and fluid equations are derived in presence of a finite coarse-graining scale: this allows a clear separation between long and short distance modes and leads to a hybrid approach where the former are treated perturbatively and the effect of the latter is encoded in external source terms for velocity, velocity dispersion, and all the higher order moments, which can be computed from N-body simulations. 

We apply the coarse-grained perturbation theory to the computation of the power spectrum and the cross-spectrum between density and velocity dispersion, and compare the results with N-body simulations, finding good agreement.
\end{abstract}

\section{Introduction}
Understanding the statistical properties of matter inhomogeneities in the universe at the percent level  is one the main goals for cosmology in the near future. A reliable description of the evolution of perturbations beyond the linear regime, i.e. at moderately high, and high wavenumbers  is indeed a very active field of investigation, since future generation of observations, such as high redshift galaxy surveys \cite{Blake:2011wn, Blake:2011rj,Eisenstein:2011sa,Bassett:2005kn,2008ASPC..399..115H,2009arXiv0912.0914L,2009arXiv0901.0721A,Percival:2009xn,Schlegel:2011wb}, are going to provide information on several aspects of the cosmological model, and with their unprecedented accuracy  call for a detailed theoretical framework to compare with. 
As an example, the location and amplitude of the Baryon Acoustic Oscillations (BAO) in the wavelength range $k \simeq 0.05 - 0.25$ $h$ Mpc$^{-1}$ are powerful probes of the expansion history of the universe and of the properties of the dark energy \cite{Eis05,2007ApJ...657...51P,2007MNRAS.381.1053P,2009MNRAS.399.1663G}. Furthermore, the yet unknown absolute neutrino mass scale will be also more efficiently constrained (or detected) by comparing data with a theoretically robust determination of the power spectrum (PS) in the high $k$ range (for a review, see \cite{Abazajian:2011dt}).

There are two approaches to deal with nonlinearities. The more established one is to use N-body simulations. However, in order to attain the required percent accuracy  on such large scales very large volumes and high resolutions are needed. The resulting limitation in computer time makes it impossible to
run simulations over the many thousands of cosmologies necessary for grid based or Markov Chain Monte Carlo (MCMC) estimation of cosmological parameters. One is therefore forced to develop interpolation methods for theoretical predictions, which limits the practical use of this approach to ``vanilla" type $\Lambda$CDM models and a restricted set of their variants, as it was discussed thoroughly in \cite{Heitmann:2008eq,Heitmann:2009cu,Lawrence:2009uk,Viel:2011bk}. 

The alternative approach is provided by Eulerian perturbation theory (PT), where the Euler-Poisson system of equations describing the self-gravitating dark matter fluid is solved perturbatively in the matter density fluctuations (for a review, see \cite{PT}, for recent applications, see for instance \cite{JK06,STT08,Wong08,SaitoII,Sefusatti:2009qh}). After its first formulation in the nineties, this framework is now experiencing a renewed interest, mainly thanks to the work of  Crocce and Scoccimarro \cite{RPTa, RPTb}, who showed that some classes of perturbative corrections can be resummed at all orders, enhancing the range of applicability of the approach towards lower redshifts and smaller scales. A number of different semi-analytic resummation methods have been proposed \cite{Valageas03,McD06,RPTa,MP07b,Matsubara07,Taruya2007,Pietroni08,TaruyaIII} and applied to the calculation of the matter PS mainly in the BAO range \cite{MP07a,MP07b,RPTBAO,Taruya:2009ir,Taruya:2010mx}. Using these methods, nonlinear effects have been computed in a variety of non-$\Lambda$CDM cosmologies, such as those with massive neutrinos \cite{LMPR09}, with primordial non-gaussianity of different shapes \cite{Bartolo:2009rb}, or with clusterizing dark energy \cite{SefuDami,Anselmi:2011ef}. A quantitative comparison between some of these approaches has been presented in \cite{Carlson:2009it}, where it was shown that in the weakly nonlinear regime, most of these approaches attain a percent level accuracy,  typically better than $~2\%$ for $k\alt 0.1$ $h$ Mpc$^{-1}$ at $z=0$.

The crucial question is now how to extend the range of applicability  of these methods towards larger wavenumbers. The obvious path is to improve the resummation schemes in order to include larger classes of perturbative corrections, not yet taken into account (for recent attempts along these lines, see for instance \cite{Bernardeau:2008fa, Anselmi:2010fs}).  However, even if all the perturbative corrections were computed or, equivalently, if the Euler-Poisson system was solved exactly, this would not imply that the dynamics of the ensemble of self-gravitating dark matter particles would be described with infinite accuracy at all scales.
Indeed, PT and all the above mentioned resummation schemes suffer from a fundamental limitation, which will strike in at small scales and late times. The point is that these approaches describe matter as a fluid, characterized in terms of density and velocity perturbations only. This is achieved by truncating the infinite tower of coupled differential equations for the moments of the particle distribution, keeping only the continuity and Euler equations, complemented with the Poisson equation relating the density field and the gravitational potential. All higher moments, in particular the velocity dispersion tensor, $\sigma^{ij}$, are neglected. This picture is fully consistent as long as there are no deviations of particle motions from a single coherent flow, the so called {\em single stream approximation}. Though for non relativistic dark matter particles  one expects the contribution of such deviations to be sub-leading, yet their inclusion leads to corrections to the PS which should be taken into account in view of future data sensitivity. This has been recently emphasized in \cite{Pueblas:2008uv}, where the contribution of $\sigma^{ij}$, which enters the Euler equation, was estimated using numerical simulations and quantified to be up to few percent for the velocity divergence PS in the BAO region at $z=0$, and slightly smaller for the density field. In a complementary analysis, Valageas \cite{Valageas:2010rx} estimated that the impact of the single stream approximation on the PS is larger than $1\,\%$ for $k\agt 0.23 \; h\,\mathrm{Mpc^{-1}}$ at $z=0$ and for $k\agt 0.44 \; h\,\mathrm{Mpc^{-1}}$ at $z=1$, therefore this assumption should be reconsidered if one aims at pushing PT-derived methods beyond such scales.

Another point which is related to the previous discussion is that to obtain a fluid description one has to take averages of the spiky 
particle distribution function over volumes typically containing many particles. In N-body simulations, this resolution scale can be identified with the grid size over which one interpolates particle positions and velocities to obtain the matter density and velocity fields according to some algorithm such as, e.g., the Cloud-in-Cell one.  In the PT framework, the averaging procedure is also instrumental in obtaining a well behaved PT expansion since, the larger the averaging volume, the smother the fluid density field. However, in general, the averaging procedure introduces a non zero velocity dispersion, even in case at the microscopic level there is no crossing of particle trajectories (see also \cite{Baumann:2010tm}). Therefore, working in the single stream approximation amounts to shrinking the averaging length to zero, and this is the reason why this scale does not appear explicitly in PT results. 

The aim of this paper is to reconsider in detail the role of the averaging procedure in PT, by keeping the averaging length explicit and defining a perturbative expansion for a coarse-grained distribution function and its related moments, including moments of the distribution up to second order (velocity dispersion). If one starts from the microscopic description and the standard kinetic equation, the smeared quantities are found to satisfy again a Vlasov equation but with a source term which contains all the information about the  behaviour of perturbations at short distances. This source term provides the way the smeared quantities are sensitive to fluctuations on scales smaller than the chosen cut-off length scale $L$. We notice in particular that the averaged velocity dispersion tensor is sourced by a contribution, due to the product of internal forces and velocity fluctuations inside the averaging volume. This is different from what happens for the analogous microscopic $\sigma^{ij}$ whose evolution is dictated by a homogenous equation, so that if one starts with vanishing initial conditions for it, there is no velocity dispersion which can be generated due to dynamics. 

At a practical level, the splitting of perturbations between long-distance modes with wavenumber $k \alt 2 \pi/L$, and small scale contributions, opens the possibility to a hybrid approach where the former are treated perturbatively while the effect of the latter, which is expected to be to some extent less sensitive to the cosmological model, can be measured from a reduced set of N-body simulations. 

The paper is organized as follows. After a brief summary of the usual microscopic nonlinear fluid equations in Section~\ref{Micro}, we define coarse-grained distribution and the corresponding moments in Section~\ref{Coarse} and deduce their evolution equations. Section~\ref{Cumulant} contains a similar analysis for the cumulant function, i.e. the moment generating functional, showing in a compact way how coarse-graining generates source terms for all higher order moments. The Fourier space version of the coarse-grained equations are then described in Sections~\ref{Fourier}, \ref{Check}, and \ref{compact}, including a first non-trivial check of the consistency of the picture.  In Section~\ref{pertsol} we develop perturbation theory for the coarse-grained moments in presence of the new source terms, which, in  Section~\ref{pertsourc}, are computed  in PT as a first approximation. In Section~\ref{sournbody}  our results for the coarse-grained PS and the cross-correlation between density and velocity dispersion are presented and compared with N-body simulations for different values of the coarse-graining scale, $L$. Finally, in Section~\ref{Conc}, we give our conclusions and discuss possible lines of developments of the approach.
\section{``Microscopic" description}
\label{Micro}

Our starting point is the {\em microscopic} distribution function for the ``elementary particles",  given by the Klimontovich density in the one-particle phase space  ($\mu$ space)
\beq
f_K(\bx,\bp,\tau) = \sum_\alpha \delta(\bx -\bx_\alpha(\tau)) \delta(\bp-\bp_\alpha(\tau))\,,
\label{Klim}
\eeq
where the sum runs over all $N$ particles. The particle's coordinate and momenta obey the Newtonian equations of motion,
\beq
\dot{\bx}_\alpha=\frac{\bp_\alpha}{a m}\,,\qquad \dot{\bp}_\alpha= - a m {\bf \nabla} \phi(\bx_\alpha, \tau)\,,
\eeq
where dots indicate derivatives w.r.t. conformal time, and the gravitational potential satisfies the Poisson equation,
\beq
\nabla^2  \phi(\bx_\alpha, \tau) = \frac{4\pi G}{a}\left[m \sum_{\beta\neq \alpha} \delta(\bx_\alpha -\bx_\beta) - \bar{\rho}  \right]\,,
\eeq
with $\bar{\rho}$ the background comoving density (which stays constant as the universe expands).

Taking moments of the Klimontovich distribution function one can define particle number density
\beq
n_K(\bx,\tau)=  \int d^3p\, f_K(\bx,\bp,\tau) =  \sum_\alpha \delta(\bx -\bx_\alpha(\tau))\,,
\eeq
and the particle current,
\beq
n_K(\bx,\tau) v^i_K(\bx,\tau) =  \int d^3p\, \frac{p^i}{a m} f_K(\bx,\bp,\tau) = \sum_\alpha \delta(\bx -\bx_\alpha(\tau))  \frac{p^i_\alpha(\tau)}{a m}\,.
\eeq
From the second order moment,
\beq
\int d^3p\, \frac{p^i}{a m}\frac{p^j}{a m} f_K(\bx,\bp,\tau)=\sum_\alpha \delta(\bx -\bx_\alpha(\tau))  \frac{p^i_\alpha(\tau)}{a m}  \frac{p^j_\alpha(\tau)}{a m}\,,
\eeq
we can extract the velocity dispersion tensor, $\sigma_K^{ij}(\bx,\tau)$,
\beqra
&&n_K(\bx,\tau) \sigma_K^{ij}(\bx,\tau) =   \int d^3p\, \left( \frac{p^i}{a m}- v^i_K(\bx,\tau)\right)  \left( \frac{p^j}{a m}- v^j_K(\bx,\tau)\right) f_K(\bx,\bp,\tau)\nonumber\\
&&\qquad\quad= \sum_\alpha \delta(\bx -\bx_\alpha(\tau))  \left( \frac{p^i_\alpha(\tau)}{a m} - v^i_K(\bx,\tau) \right)   \left( \frac{p^j_\alpha(\tau)}{a m} - v^j_K(\bx,\tau) \right) \,,
\label{sigmic}
\eeqra
which is non-vanishing only in those points were two or more particle trajectories cross (shell-crossing).

The conservation of the density (\ref{Klim})  along particle trajectories in phase space gives the dynamical equation
\beq
\left[\frac{\partial\,}{\partial \tau} + \frac{p^i}{ma} \frac{\partial\,}{\partial x^i} - a m \nabla^i \phi(\bx) \frac{\partial\,}{\partial p^i}\right] f_K(\bx,\bp,\tau)=0\,,
\label{Vlasov}
\eeq
known as the Klimontovich equation.

\section{From particles to fluid}
\label{Coarse}

In any practical use in cosmology we are interested in scales much larger than the mean inter-particle distance. Therefore, the relevant distribution function is obtained by volume averaging over the resolution scale,
\beq
\bar{f}(\bx,\bp,\tau)\equiv \frac{1}{V}\int d^3y\, {\cal W}\left(\left|\frac{\by}{L}\right|\right) f_K(\bx+\by,\bp,\tau)\,,
\label{fbar}
\eeq
where $ {\cal W} (|z|) $ is a window function normalized to unity
\beq
 \frac{1}{V}\int d^3y\, {\cal W}\left(\left|\frac{\by}{L}\right|\right)=1\,,
\eeq
 and rapidly vanishing for $|z|>1$. If a volume $V\sim L^3$ typically contains many elementary particles, the distribution $\fb$ is a smooth one, compared to the spiky Klimontovich one. The momentum integral of $\fb$ gives the smoothed number  density of particles,
\beq
\nb(\bx,\tau) = \int d^3p\, \fb(\bx,\bp,\tau) =\frac{1}{V}\sum_\alpha  {\cal W}\left(\left|\frac{\bx-\bx_\alpha}{L}\right|\right)\,.
\label{nmic}
\eeq
If the window function $ {\cal W}(|z|)$ is the {\it top hat} one ({\it i.e.} $ {\cal W}(|z|)=1$ for $|z|\le 1$, and  $ {\cal W}(|z|)=0$ for $|z|>1$), then the sum in (\ref{nmic}) counts the particles contained in   a volume $V$ centered around $\bx$ and $\nb(\bx,\tau)$ is the corresponding number density.

The first moment of the distribution function defines the peculiar velocity field,
\beqra
\vb^i(\bx,\tau) &=& \frac{1}{{\nb}(\bx,\tau)}  \int d^3p\, \frac{p^i}{a\,m}\fb(\bx,\bp,\tau)\nonumber\\
&=&  \frac{1}{{\nb}(\bx,\tau)}  \frac{1}{V}\sum_\alpha \frac{p_\alpha^i}{a\,m} {\cal W}\left(\left|\frac{\bx-\bx_\alpha}{L}\right|\right)\,.
\label{vav}
\eeqra
By taking again a top hat filter, ${\vb}^i(\bx,\tau)$ is given by the velocity of the center of mass (c.o.m.) of the particles contained in $V$.
The second moment
\beq
\frac{1}{{\nb}(\bx,\tau)}  \int d^3p\, \frac{p^i}{a\,m}\frac{p^j}{a\,m}{\fb}(\bx,\bp,\tau),
\eeq
gives
\beq
\frac{1}{{\nb}(\bx,\tau)}  \frac{1}{V}\sum_\alpha \frac{p_\alpha^i}{a\,m}  \frac{p_\alpha^j}{a\,m} {\cal W}\left(\left|\frac{\bx-\bx_\alpha}{L}\right|\right) = {\vb}^i(\bx,\tau) {\vb}^j(\bx,\tau) +\bar{\sigma}^{ij}(\bx,\tau)\,,
\label{2mom}
\eeq
where $ \bar{\sigma}^{ij}(\bx,\tau)$ is the velocity dispersion of the particles in $V$ around the c.o.m. value $\bar{v}^i$,
\beq
\bar{\sigma}^{ij}(\bx,\tau) =   \frac{1}{{\nb}(\bx,\tau)} \frac{1}{V}\sum_\alpha \delta v^i_\alpha \, \delta v^j_\alpha \, {\cal W}\left(\left|\frac{\bx-\bx_\alpha}{L}\right|\right)\,,
\label{sigbar0}
\eeq
with 
\beq
 \delta v^i_\alpha\equiv \frac{p^i_\alpha}{a\,m} - {\vb}^i\,.
 \eeq

While the coarse-grained density is  just the volume average of the microscopic one,
\beq
\nb(\bx,\tau) =  \frac{1}{V}\int d^3y\, {\cal W}\left(\left|\frac{\by}{L}\right|\right) n_K(\bx+\by,\tau)\,,
\label{nbarra}
\eeq
the coarse-grained velocity is given by the volume average weighted by the microscopic density field, and it is therefore different from the filtered density field,
\beqra
 \vb^i(\bx,\tau)& =&  \frac{1}{ \nb(\bx,\tau)}\frac{1}{ V }\int d^3y\, {\cal W}\left(\left|\frac{\by}{L}\right|\right) n_K(\bx+\by,\tau) v^i_K(\bx+\by,\tau)\nonumber\\
&\neq& \frac{1}{ V }\int d^3y\, {\cal W}\left(\left|\frac{\by}{L}\right|\right) v^i_K(\bx+\by,\tau)\,.
\label{vbari}
\eeqra
Moreover, the velocity dispersion obtained from the second moment of $\fb$, $\bar{\sigma}^{ij}$ is given by two contributions, 
\beqra
&& \bar{\sigma}^{ij}(\bx,\tau) =  \frac{1}{ \nb(\bx,\tau)}\frac{1}{ V }\int d^3y\, {\cal W}\left(\left|\frac{\by}{L}\right|\right) n_K(\bx+\by,\tau) \sigma_K^{ij}(\bx+\by,\tau)\nonumber\\
&&+   \frac{1}{ \nb(\bx,\tau)}\frac{1}{ V }\int d^3y\, {\cal W}\left(\left|\frac{\by}{L}\right|\right) n_K(\bx+\by,\tau)\delta v^i(\bx+\by,\tau)\delta v^j(\bx+\by,\tau) \,,
\label{sigbar}
\eeqra
with
\beq
\delta v^i(\bx+\by,\tau) \equiv v^i_K(\bx+\by,\tau) -\vb^i(\bx,\tau)\,,
\eeq
where the first contribution is the microscopic one coming from the crossing of particle trajectories  (see eq.~(\ref{sigmic})), while the second one is given by the dispersion of the velocities of all the particles contained in the coarse-graining volume, and can be non-zero also in absence of microscopic shell-crossing. If one would iterate the coarse-graining procedure, by averaging $\bar{f}$ over a larger volume, e.g. $(2L)^3$, the new velocity dispersion would be given by an expression analogous to eq.~(\ref{sigbar}), where the role of the microscopic velocity dispersion $\sigma_K^{ij}$ would now be played by $\bar{\sigma}^{ij}$ and the second term would be given by the velocity dispersion among the c.o.m. velocities of the $L^3$ volumes contained in the $(2L)^3$ one. According to this point of view, the coincidence between non-vanishing velocity dispersion and shell crossing is exact only in the $L\to 0$ limit, and becomes more and more irrelevant as $L$ grows beyond the typical interparticle distance. Indeed, as we will see in eq.~(\ref{newsig}), the macroscopic velocity dispersion is a dynamical quantity, which is generated by the product of internal forces and velocity fluctuations inside the averaging volume.

Applying (\ref{Vlasov}) on (\ref{fbar}) we derive the equation 
\beqra
&&\left[\frac{\partial\,}{\partial \tau} + \frac{p^i}{ma} \frac{\partial\,}{\partial x^i} - a m \nabla^i_x \bar{\phi}(\bx,\tau) \frac{\partial\,}{\partial p^i}\right] \bar{f}(\bx,\bp,\tau) \nonumber\\
&&\quad = \frac{a m}{V}\int d^3y \, {\cal W}\left(\left|\frac{\by}{L}\right|\right) \nabla_{x+y}^i \delta\phi(\bx+\by,\tau)\, \frac{\partial\,}{\partial p^i}  \delta f (\bx+\by,\bp,\tau)\,,
\label{vlbar}
\eeqra
where we have split the gravitational potential $\phi$ into a long-wavelength and a short-wavelength part,
\beq
\phi(\bx+\by,\tau) = \bar{\phi}(\bx,\tau) + \delta\phi(\bx+ \by,\tau)\,,
\eeq
with
\beq
 \bar{\phi}(\bx,\tau ) \equiv \frac{1}{V}\int d^3y\, {\cal W}\left(\left|\frac{\by}{L}\right|\right) \phi(\bx+\by,\tau) \,,
\label{phibar}
\eeq
and, moreover, we have defined
\beq
 \delta f (\bx+\by,\bp,\tau) \equiv  f_K(\bx+\by,\bp,\tau) -  \fb(\bx,\bp,\tau)\,.
\eeq
The term at the r.h.s. of eq.~(\ref{vlbar}) represents the contribution of the short-wavelength fluctuations to the evolution of the long-wavelength ones. Neglecting it, one obtains the Vlasov equation, which is the starting point for cosmological perturbation theory (see for instance \cite{PT}).  In other terms, PT and the resummation methods considered so far assume the $L\to 0$ limit. In this paper, we will work instead with the complete equation, and discuss the effect of the r.h.s. on the evolution of the coarse-grained quantities derived from $\fb$.

Taking moments of eq.~(\ref{vlbar}) we obtain equations for the coarse-grained density, velocity,  velocity dispersion, and so on. To keep the notation as compact as possible, in the following  we will omit the time-dependence, where obvious.
The continuity equation, as expected, is not modified, 
\beq
\frac{\partial\,}{\partial \tau} \bar{n}(\bx) + \frac{\partial\,}{\partial x^i} \left( \bar{n}(\bx) \bar{v}^i(\bx)\right)=0\,.
\label{newcon}
\eeq
On the other hand, the equation for the average velocity reads,
\beqra
&& \frac{\partial\,}{\partial \tau}  \vb^i(\bx) +  {\cal H}\vb^i(\bx)  +\vb^k(\bx)   \frac{\partial\,}{\partial x^k} \vb^i(\bx) + \frac{1}{\nb(\bx)}  \frac{\partial\,}{\partial x^k} (\nb(\bx) \bar{\sigma}^{ki}(\bx))  \nonumber \\
&&= -  \nabla_x^i \bar{\phi}(\bx)- \frac{1}{V}  \int d^3y \;{\cal W}\left(\left|\frac{\by}{L}\right|\right)   \frac{n(\bx+\by)}{\nb(\bx)}\,\nabla^i_{x+y} \delta\phi(\bx+\by)\,.
\label{neweul}
\eeqra
Compared to the  Euler equation, it has two extra terms. The last one at the l.h.s. describes the effect of the velocity dispersion around the mean velocity, whereas the last one at the r.h.s. accounts for the distribution of matter inside the coarse-graining volume $V$. In other words, while  $ -\nabla_x^i \bar{\phi}(\bx)$ represents the gravitational force generated by the monopole of the mass distribution in $V$, the second term at the r.h.s. is the contribution to the force of the higher mass multipoles. 

Eq.~\ref{neweul} was derived in \cite{Buchert:2005xj}, where, in order to close the system, different dynamical models were considered to compute $\bar\sigma^{ij}$ and the source term at the RHS. On the other hand, in ref.~\cite{Pueblas:2008uv}, the $\bar{\sigma}^{ij}$ tensor was measured from a simulation, whereas the source term was not considered. In this paper we take a different approach, by adding the dynamical equation for the long-distance quantity $\bar\sigma^{ij}$ and by considering different approximations, namely PT and N-body simulations, to deal with the short-distance information encoded in the source terms.

Taking the second moment of eq.~(\ref{vlbar}) we get the equation for the macroscopic velocity dispersion,
\beqra
&&  \frac{\partial\,}{\partial \tau} \bar{\sigma}^{ij} + 2  {\cal H} \bar{\sigma}^{ij}   +\vb^k  \frac{\partial\,}{\partial x^k}   \bar{\sigma}^{ij}+
 \bar{\sigma}^{ik}  \frac{\partial\,}{\partial x^k}  \vb^j + \bar{\sigma}^{jk}  \frac{\partial\,}{\partial x^k}  \vb^i+
\frac{1}{\nb}   \frac{\partial\,}{\partial x^k} \left(\nb\, \bar\omega^{ijk} \right) \nonumber\\
&&= -\frac{1}{V}  \int d^3y \; \;{\cal W}\left(\left|\frac{\by}{L}\right|\right)  \frac{n(\bx+\by)}{\nb(\bx)} \nonumber\\
&&\qquad\qquad\qquad\qquad\times \left[\delta v^j(\bx+\by) \nabla^i_{x+y} +\delta v^i(\bx+\by) \nabla^j_{x+y} 
  \right] \delta\phi(\bx+\by)\,.\nonumber\\
  &&
  \label{newsig}
\eeqra
The equations for the irreducible third moment, $ \bar\omega^{ijk}$, as those for all the higher order ones, can be obtained in a completely analogous way, see also next section for a compact treatment of all higher order moments.

The term at the r.h.s. of the equation above provides a source for the velocity dispersion, which is sourceless if one starts from the Vlasov equation instead than from eq.~(\ref{vlbar}). It clarifies the ``microscopic" origin of velocity dispersions, which, as anticipated, are generated by the product of internal forces and velocity fluctuations inside the averaging volume.

Notice that the smeared gravitational potential $\bar{\phi}(\bx,\tau)$ is related to the coarse-grained density fluctuation,  $\bar\delta(\bx)\equiv\ m \bar{n}(\bx)/\bar{\rho}-1$ by the Poisson equation
\beq 
\nabla^2 \bar{\phi}(\bx,\tau) = \frac{3}{2} {\cal H}^2 \Omega_m \bar\delta(\bx) \,,
\eeq
while $ \delta\phi(\bx,\tau)$ satisfies the same equation with the density fluctuations corresponding to $\delta n(\bx)$ at the r.h.s., with
\beq \delta n(\bx) = \int d^3 p\, \delta f(\bx,\bp) \,.
\eeq

As a final remark we take into account the averaged vorticity field, $\bar\bw=\nabla \times \bf \vb$. It also receives a non-vanishing source term, given by
\beq
 -\frac{1}{\bar{n}(\bx)} \nabla \bar{n}(\bx) \times \bf{J_v} = -\nabla \bar{\delta}\times \bf{J_v} +\cdots \,,
 \label{sourcew}
 \eeq
 where $\bf{J_v}$ is the second term at the RHS of eq.~(\ref{neweul}), and it is also turned on by the $\sigma^{ki}$-dependent term at the LHS. The source term in eq.~(\ref{sourcew}) is suppressed by the gradient of the coarse-grained density fluctuation $\bar{\delta}$ with respect to the source terms for the velocity divergence and for the velocity dispersion. Moreover, the effect of the  $\sigma^{ki}$-dependent term was shown to have a subleading impact on the density and velocity PS's in \cite{Pueblas:2008uv}. Therefore we will neglect vorticity in the rest of this paper, leaving it for a future analysis.

\section{Coarse-grained cumulant}
\label{Cumulant}

Before proceeding to computations, we discuss the generalization of eqs.~(\ref{neweul}) and (\ref{newsig}) to moments of arbitrarily high order.  It is useful to consider the Fourier transforms of the distribution functions w.r.t. momentum:
\beq
M(\bx,\bl,\tau) = \int d^3 p\, e^{i\frac{\bl\cdot\bp}{ma}} \,f(\bx,\bp,\tau).
\label{M}
\eeq
The microscopic distribution function $f(\bx,\bp,\tau)$ can be thought as the Klimontovich one, eq.~(\ref{Klim}), or a smoothed version of it, defined as in eq.~(\ref{fbar}) but taking the smoothing scale $L=L_{UV}$, where $L_{UV}$ is much smaller than any cosmological scale we are interested in. In practice, we are assuming that $f$ satisfies the standard Vlasov equation, that is, eq.~(\ref{vlbar}) with vanishing RHS.
$M(\bx,\bl,\tau)$ is the moment generating functional, since its derivatives w.r.t. $\bl$, evaluated at $\bl=0$, give
\beqra
M(\bx,\bl=0,\tau)&&= n(\bx,\tau)\,,\nonumber\\
 -i\left.\frac{\partial M(\bx,\bl,\tau)}{\partial l_j}\right|_{\bl=0} &&= n(\bx,\tau) v^j(\bx,\tau)\,,\nonumber\\
  (-i)^2\left.\frac{\partial^2 M(\bx,\bl,\tau)}{\partial l_j\partial l_k}\right|_{\bl=0} &&= n(\bx,\tau) (v^j(\bx,\tau) v^k(\bx,\tau) + \sigma^{jk}(\bx,\tau) )\,,\nonumber\\
\ldots&&
\eeqra
and so on.
Taking the logarithm of $M(\bx,\bl,\tau)$ we get the cumulant generating functional, $C(\bx,\bl,\tau) = \log M(\bx,\bl,\tau)$, whose derivatives with respect to $\bl$ give the irreducible moments, 
\beqra
C(\bx,\bl=0,\tau)&&= \log \left[ n(\bx,\tau)\right] \,, \nonumber\\
 -i\left.\frac{\partial C (\bx,\bl,\tau)}{\partial l_j}\right|_{\bl=0} &&=  v^j(\bx,\tau)\,,\nonumber\\
  (-i)^2\left.\frac{\partial^2 C(\bx,\bl,\tau)}{\partial l_j\partial l_k}\right|_{\bl=0} &&= \sigma^{jk}(\bx,\tau) \,,\nonumber\\
\ldots&&
\eeqra
and so on.

By single stream regime one refers to the situation where the fluid momentum is single valued at each point in space, {\it i.e.} particle trajectories do not cross. In this regime, the distribution function takes the form
\beq
f(\bx,\bk,\tau)= g(\bx,\tau) \,\delta_D(\bk-{\bf P}(\bx,\tau))\,.
\eeq
Inserting this form in~(\ref{M}), we can derive the expression for $M$, and then for $C$ in the single stream regime
\beq
C(\bx,\bl,\tau) = i\, \frac{\bl\cdot {\bf P}(\bx,\tau) }{a m} + \log  g(\bx,\tau)\,,
\label{Css}
\eeq
from which we can verify that, being $C$ linear in $\bl$, all cumulants of order higher than one vanish.

Assuming that the microscopic theory satisfies the Vlasov equation, we can work out the corresponding equations for $M(\bx,\bl,\tau)$ and $C(\bx,\bl,\tau)$. The equation for the latter is
\beqra
&&\left[\frac{\partial\,}{\partial \tau} + {\cal H}\,  l^j \frac{\partial\,}{\partial l^j}  -i \frac{\partial^2\;}{\partial l^j\partial x^j }\right] C(\bx,\bl,\tau) -i \frac{\partial C(\bx,\bl,\tau)}{\partial l^j} \frac{\partial C(\bx,\bl,\tau)}{\partial x^j}\nonumber\\
&&\qquad\qquad\qquad\qquad\qquad\qquad \qquad\qquad\qquad = -i\, l^j \nabla^j \phi(\bx,\tau)\,,
\label{vlasovC}
\eeqra
where we recognize the term responsible for nonlinear couplings in the last term at the l.h.s., and the source term at the r.h.s.
Since the source term is linear in $\bl$, it only appears in the equation for the velocity (the Euler equation). Moreover, the structure of the equation -- and in particular  of the nonlinear term -- ensures that the form (\ref{Css}) is a fixed point, that is, nonlinear terms in $\bl$, corresponding to velocity-dispersion and higher order moments, are not generated if they are not present initially.

Starting from the coarse-grained distribution function $\bar{f}(\bx,\bp,\tau)$, see eq.~(\ref{fbar}), we can define a coarse-grained moment generating functional,
\beq \bar{M}(\bx,\bl,\tau) =  \int d^3 p\, e^{i\frac{\bl\cdot\bp}{ma}} \,\bar{f}(\bx,\bp,\tau)= \frac{1}{V}\int d^3y\, {\cal W}\left(\left|\frac{\by}{L}\right|\right) M(\bx+\by,\bl,\tau)\,,
\label{newM}
\eeq
whose derivatives give $\bar{n}$, $\bar{n} \bar{v}^j$,  $\bar{n} (\bar{v}^j\bar{v}^k +\bar{\sigma}^{jk})$, and so on. The relation between the cumulants generating functionals is therefore
\beq
 \bar{C}(\bx,\bl,\tau) = \log\left[  \frac{1}{V}\int d^3y\, {\cal W}\left(\left|\frac{\by}{L}\right|\right)  \exp[C(\bx+\by,\bl,\tau)]  \right]\,.
 \label{Cc}
\eeq
From the derivatives of $\bar{C}(\bx,\bl,\tau)$ we get the averaged cumulants:
\beqra
&&\bar{C}(\bx,\bl=0,\tau)= \log \left[ \bar{n}(\bx,\tau)\right]    \,, \nonumber\\
&& -i\left.\frac{\partial \bar{C} (\bx,\bl,\tau)}{\partial l_j}\right|_{\bl=0} =  \bar{v}^j(\bx,\tau)  \,,\nonumber\\
  && (-i)^2\left.\frac{\partial^2 \bar{C}(\bx,\bl,\tau)}{\partial l_j\partial l_k}\right|_{\bl=0} = \sigmab^{jk}(\bx,\tau)\,,\nonumber\\
  &&\qquad\qquad \cdots\,.
\eeqra
From eq.~(\ref{Css}) and (\ref{Cc}) we can determine the expression for the coarse-grained cumulants generating functional when the microscopic theory is in the single stream regime,
\beqra
&&\bar{C}(\bx,\bl,\tau)=\nonumber\\
&&  \log \left[ \frac{1}{V}\int d^3y\, {\cal W}\left(\left|\frac{\by}{L}\right|\right) \,g(\bx+\by,\tau) \, \exp\left(i\frac{\bl\cdot {\bf P}(\bx+\by,\tau)}{ma}\right) \right]\,,
\eeqra
from which we see that, if the particle momentum $P(\bx,\tau)$ is not uniform inside the coarse-graining volume $V$, all orders in $\bl$ are present in $\bar{C}$, and therefore all coarse-grained cumulants are non-vanishing, despite the microscopic theory being in the single stream regime.

The evolution equation for $\bar{C}$ can be derived using that for $C$, eq.(\ref{vlasovC}),
\beqra
&&\left[\frac{\partial\,}{\partial \tau} + {\cal H}\,  l^j \frac{\partial\,}{\partial l^j}  -i \frac{\partial^2\;}{\partial l^j\partial x^j }\right] \bar{C}(\bx,\bl,\tau) -i \frac{\partial \bar{C}(\bx,\bl,\tau)}{\partial l^j} \frac{\partial \bar{C}(\bx,\bl,\tau)}{\partial x^j}\nonumber\\
&& = -i\, l^j \left(\nabla^j \bar{\phi}(\bx,\tau) + \frac{1}{V}\int d^3y\, {\cal W}\left(\left|\frac{\by}{L}\right|\right) \,e^{\delta C (\bx+\by,\bl,\tau)}\nabla^j_{\bx+\by} \delta\phi(\bx+\by,\tau)  \right) \,,\nonumber\\
&&
\label{vlasovCb}
\eeqra
where 
\beq
\delta C(\bx+\by,\bl,\tau) \equiv C(\bx+\by,\bl,\tau) -\bar{C}(\bx,\bl,\tau) \,.
\eeq
One can check that taking derivatives of  eq.~(\ref{vlasovCb}) w.r.t. $\bl$ gives eqs.~(\ref{newcon}), (\ref{neweul}), (\ref{newsig}), and so on.

The second term at the r.h.s. of eq.~(\ref{vlasovCb}) contains all orders in $\bl$ even when the microscopic theory is in the single stream regime --provided $g(\bx,\tau)$ and $P(\bx,\tau)$ in eq.~(\ref{Css}) are not constant in the coarse-graining volume $V$ -- and therefore it represents a source term for all cumulants of order greater or equal  than one. Notice that, unlike eq.~(\ref{vlasovC}), the equation for $\bar{C}$ is not closed, since it requires input from the microscopic dynamics encoded in $\delta C$ and $\delta \phi$  appearing in the source term. The equation for  $\delta C$ can be obtained by subtracting eq.~(\ref{vlasovCb}) from eq.~(\ref{vlasovC}),
\beqra
&&\left[\frac{\partial\,}{\partial \tau} + {\cal H}\,  l^j \frac{\partial\,}{\partial l^j}  -i \frac{\partial^2\;}{\partial l^j\partial x^j }\right] \delta C(\bx,\bl,\tau) -i \frac{\partial\delta C(\bx,\bl,\tau) }{\partial l^j} \frac{\partial \delta C(\bx,\bl,\tau) }{\partial x^j}\nonumber\\
&& \qquad\qquad\quad-i \frac{\partial\delta C(\bx,\bl,\tau) }{\partial l^j} \frac{\partial \bar{C}(\bx,\bl,\tau) }{\partial x^j}-i \frac{\partial \bar{C}(\bx,\bl,\tau) }{\partial l^j} \frac{\partial \delta C(\bx,\bl,\tau) }{\partial x^j}\nonumber\\
&& = -i\, l^j \left(\nabla^j \delta\phi(\bx,\tau) - \frac{1}{V}\int d^3y\, {\cal W}\left(\left|\frac{\by}{L}\right|\right) \,e^{\delta C (\bx+\by,\bl,\tau)}\nabla^j_{\bx+\by} \delta\phi(\bx+\by,\tau)  \right) \,,\nonumber\\
&&
\label{vlasovdeltaC}
\eeqra
and, of course, the two equations (\ref{vlasovCb}) and (\ref{vlasovdeltaC}), together with the Poisson equations for $\bar{\phi}$ and $\delta \phi$, form a closed system describing the same physics as eq.~(\ref{vlasovC}) (and the Poisson equation for $\phi$). However, the splitting of $C$ in a coarse-grained part, $\bar{C}$, and a fluctuation $\delta C$ suggests to treat them differently, {\it e.g.} by using perturbation theory for the former and measuring the latter from $N$-body simulations.

\section{Fourier space}
\label{Fourier}

In the following, we will deal with the system of equations~(\ref{newcon}), (\ref{neweul}), (\ref{newsig}), where we will set $\omega_{ijk}=0$. It is convenient to Fourier transform the equations.

The equations for the coarse-grained density fluctuation ({\it i.e.} the Fourier transform of $\bar\delta(\bx)$, where $n_0$ is the average comoving number density), the velocity divergence ($\bar\theta(\bx)=\nabla^i \bar{v}^i$) and velocity dispersion are
\beqra
&&\dot{\bar\delta}(\bk)+\bar\theta(\bk) +\int d^3q_1 d^3q_2 \,\delta_D(\bk-\bq_1-\bq_2)\, \frac{\bk\cdot\bq_2}{q_2^2}\, \bar\delta(\bq_1)\bar\theta(\bq_2)=0\,,\label{deltadot}\\
&&\dot{\bar\theta}(\bk)+{\cal H}\, \bar\theta(\bk)+ \frac{3}{2}{\cal H}^2\, \Omega_m \bar\delta(\bk) -  k_i k_j \bar\sigma^{ij}(\bk)  =  -J_\theta(\bk) \nonumber\\
&&-\int d^3q_1 d^3q_2 \,\delta_D(\bk-\bq_1-\bq_2)\, \left[ \frac{k^2\, \bq_1\cdot\bq_2}{2 \,q_1^2 q_2^2}\, \bar\theta(\bq_1)\bar\theta(\bq_2)
- k_i q_{2j} \bar\sigma^{ij}(\bq_1) \bar\delta(\bq_2)
\right]\,,\nonumber\\
&& \label{thetadot}\\
%&&\qquad\quad\left.+ \frac{3}{2}{\cal H}^2\, \Omega_m \,\frac{1}{2}\left(\frac{\bk\cdot\bq_1}{q_1^2}+\frac{\bk\cdot\bq_2}{q_2^2}\right) \delta_>(\bq_1)\delta_>(\bq_2)\right]  %\nonumber\\
&&\nonumber\\
&& \dot{\bar\sigma^{ij}}(\bk)+2\,{\cal H}\, \bar\sigma^{ij}(\bk) =-\int d^3q_1 d^3q_2 \,\delta_D(\bk-\bq_1-\bq_2)\, \left[\frac{\bq_1\cdot\bq_2}{q_1^2} \bar\theta(\bq_1)\bar\sigma^{ij}(\bq_2)\right.\nonumber\\
&& \qquad\qquad\qquad\qquad\quad\left.+\frac{ q_{1k}}{q_1^2}\bar\theta(\bq_1)\left(q_1^i\bar\sigma^{kj}(\bq_2)+q_1^j\bar\sigma^{ki}(\bq_2)\right) \right] - J_\sigma^{ij}(\bk)\,,\nonumber\\
%-\frac{3}{2} {\cal H}^2\,\Omega_m \frac{q_1^iq_2^j+q_1^jq_2^i}{q_1^2 q_2^2} \theta_>(\bq_1)\delta_>(\bq_2)
%\right]\,\nonumber\\
%&&+\frac{3}{2} {\cal H}^2\Omega_m \int d^3q_1 d^3q_2 d^3q_3 \delta_D(\bk-\bq_1-\bq_2-\bq_3) \frac{q_2^iq_3^j+q_2^jq_3^i}{q_2^2 q_3^2}\delta_>(\bq_1) \theta_>%(\bq_2)\delta_>(\bq_3)\,,\nonumber\\
&&
\label{sigmadot}
\eeqra
where we have neglected vorticity, {\it i.e.}, we have expressed velocity in terms of its divergence, $v^j(\bp)=-i p^j/p^2\, \theta(\bp)$. In the equation for the velocity we have also neglected terms of order $ \bar \sigma^{ij}\, \bar \delta\,\partial_j \bar\delta$ and higher coming from the last term at the l.h.s. of eq.~(\ref{neweul}). This can be motivated since the non-linearity of the coarse-grained quantities $\bar{\delta}$, $\bar\theta$, $\bar{\sigma}^{ij}$ can be reduced at will by taking a large enough averaging scale, $L$.

The source terms, $-J_\theta(\bk)$ and $-J_\sigma^{ij}(\bk)$, are given by the Fourier transform of the divergence of the second term at the r.h.s. of eq.~(\ref{neweul}) and of the r.h.s. of eq.~(\ref{newsig}), respectively.

The symmetric tensor $\bar{\sigma}^{ij}(\bk)$ can be decomposed as
\beq
\bar\sigma^{ij}(\bk) = \frac{1}{2}\left(\delta^{ij}-\frac{k^ik^j}{k^2}\right) \bar\sigma(\bk) + \frac{3}{2}\left(\frac{k^ik^j}{k^2}-\frac{1}{3}\delta^{ij}\right) \bar\Sigma(\bk) +\bar h^{ij}(\bk)\,,
\eeq
where 
\beq
\bar \sigma(\bk)=Tr(\bar\sigma(\bk))\,,\;\;\;\;\bar\Sigma(\bk)=\frac{k_ik_j}{k^2}\bar\sigma^{ij}(\bk)\,,\;\;\; Tr(h(\bk))=k_i \bar h^{ij}(\bk)=0\,.
\eeq
Neglecting the traceless and transverse tensor $\bar h^{ij}$ we can derive, from (\ref{sigmadot}), the equations for $\bar \sigma$ and $\bar \Sigma$,
\beqra
&& \dot{\bar\sigma}(\bk)+2\,{\cal H}\, \bar \sigma(\bk) =-\int d^3q_1 d^3q_2 \,\delta_D(\bk-\bq_1-\bq_2)\nonumber\\
&&\,\times \left[\left(1+\frac{\bq_1\cdot\bq_2}{q_1^2} -  \frac{(\bq_1\cdot\bq_2)^2}{q_1^2q_2^2} \right) \bar \theta(\bq_1)\bar\sigma(\bq_2)
 + \,3\left( \frac{(\bq_1\cdot\bq_2)^2}{q_1^2q_2^2} -\frac{1}{3} \right)\bar \theta(\bq_1)\bar\Sigma(\bq_2)\right]
\nonumber\\
&& \qquad\qquad\qquad\qquad\quad-  J_{\sigma}(\bk)\,,
%+ \,3\left( \frac{(\bq_1\cdot\bq_2)^2}{q_1^2q_2^2} -\frac{1}{3} \right)\theta_<(\bq_1)\Sigma_<(\bq_2) -3 {\cal H}^2\Omega_m\frac{\bq_1\cdot\bq_2}{q_1^2q_2^2}\theta_>%(\bq_1)\delta_>(\bq_2)
%\right]\,,\nonumber\\
%&&+3 {\cal H}^2\Omega_m \int d^3q_1 d^3q_2 d^3q_3 \delta_D(\bk-\bq_1-\bq_2-\bq_3) \frac{\bq_2 \cdot \bq_3}{q_2^2 q_3^2}\delta_>(\bq_1) \theta_>(\bq_2)\delta_>%(\bq_3)\,,
\label{sigmaT}\\
&& \dot{\bar\Sigma}(\bk)+2\,{\cal H}\, \bar\Sigma(\bk) =-\int d^3q_1 d^3q_2 \,\delta_D(\bk-\bq_1-\bq_2) \nonumber\\
&&\times\left\{\left[\frac{\bq_1\cdot\bq_2}{2\,q_1^2}\left(1-\frac{(\bk\cdot\bq_2)^2}{k^2q_2^2}\right)+\frac{(\bk\cdot\bq_1)^2}{k^2q_1^2}-\frac{\bk\cdot\bq_1\,\bk\cdot\bq_2\,\bq_1\cdot\bq_2}{k^2q_1^2q_2^2}
\right]\bar\theta(\bq_1)\bar\sigma(\bq_2)
\right.\nonumber\\
&&+\left.\left[  \frac{\bq_1\cdot\bq_2}{2\,q_1^2}\left(3\frac{(\bk\cdot\bq_2)^2}{k^2q_2^2}-1\right) +3\frac{\bk\cdot\bq_1\,\bk\cdot\bq_2\,\bq_1\cdot\bq_2}{k^2q_1^2q_2^2}-\frac{(\bk\cdot\bq_1)^2}{k^2q_1^2} \right]\bar\theta(\bq_1)\bar\Sigma(\bq_2) \right\}\nonumber\\
%&&\left.-3{\cal H}^2\Omega_m \frac{\bk\cdot\bq_1\,\bk\cdot\bq_2}{k^2q_1^2q_2^2} \theta_>(\bq_1)\delta_>(\bq_2)\right\}\,\nonumber\\
%&& +3 {\cal H}^2\Omega_m \int d^3q_1 d^3q_2 d^3q_3 \delta_D(\bk-\bq_1-\bq_2-\bq_3) \frac{\bk \cdot \bq_2\,\bk \cdot \bq_3}{k^2 q_2^2 q_3^2}\delta_>(\bq_1) \theta_>%(\bq_2)\delta_>(\bq_3)\,,\nonumber\\
&& \qquad\qquad\qquad\quad-  J_{\Sigma}(\bk)\,,
 \label{Sigma}
\eeqra
where
\beq
J_{\sigma}(\bk)=Tr(J_\sigma(\bk))\,,\;\;\;\;J_\Sigma(\bk)=\frac{k_ik_j}{k^2}J^{ij}_\sigma(\bk)\,.
\eeq

\section{A first check}
\label{Check}

As a first application of the above equations, we can compute the zero-mode ({\it i.e.} the volume average in physical space)  of the trace of the velocity dispersion. Taking the average in eq.~(\ref{sigmaT}) we get
\beq
\partial_\lambda\,\bar\sigma_0(\lambda)+2\,\bar\sigma_0(\lambda) =\int d^3q\left[ -P_{\theta \sigma}(q) + 2\, P_{\theta \Sigma}(q)\right]  -\frac{j^\sigma_0(\lambda)}{{\cal H}}\, ,
\label{sigav}
\eeq
where we have switched from $\tau$  to 
\beq\lambda\equiv \log a/a_{in}\,,
\label{lambda}
\eeq
and we have defined the volume averages (zero modes) of $\bar\sigma$ and $J_\sigma$
\beq
\delta_D(\bk) \, \bar\sigma_0(\lambda)\equiv \langle \bar \sigma(\bk)\rangle\qquad\qquad\qquad \delta_D(\bk) \, j^\sigma_0(\lambda)\equiv \langle J_\sigma(\bk) \rangle\,.
\eeq
Notice that the brackets indicate a spatial average over the {\em total volume}, $V_{tot} = (2 \pi)^3 \delta_D(\bk)$, not to be confused with the volume $V \sim L^3$ over  which we define the coarse-grained quantities.
The power spectra in eq.~(\ref{sigav}) are defined as
\beqra
&&\langle- \frac{\bar \theta(\bq_1)}{{\cal H}}\, \bar\sigma(\bq_2)\rangle \equiv \delta_D(\bq_1+\bq_2) P_{\theta \sigma}(q_1)\,,\label{cts}\\
&&\langle -\frac{\bar \theta(\bq_1)}{{\cal H}} \,\bar\Sigma(\bq_2)\rangle \equiv \delta_D(\bq_1+\bq_2) P_{\theta \Sigma}(q_1)\,.\label{ctS}
\eeqra
Consider now the r.h.s. of eq.~(\ref{sigav}) in two opposite limits. If one takes the coarse-graining scale $L$ to zero, the source term does not contribute, as it encodes physics inside the coarse-graining volume, which shrinks to zero. The remaining contributions are given by the cross-correlators between the peculiar velocity field and the velocity dispersions $\bar\sigma$ and $\bar \Sigma$, eqs.~(\ref{cts}) and (\ref{ctS}). However, in the same limit $L \rightarrow 0$, the latter receive non-vanishing contributions only from the microscopic dispersion velocity defined in eq.~(\ref{sigmic}), {\it i.e.}, only from departures from the single stream regime at small scales due to the crossing of particle trajectories (see also eq.~(\ref{sigbar})). This regime is out of reach for perturbation theory and semi-analytic resummation methods, and therefore one should resort to numerical simulations in order to compute these contributions.

In the opposite limit, {\it i.e.} when $L^3\sim V \to V_{tot}\to \infty$, the situation is just the opposite. The contribution of  the integral at the r.h.s. of eq.~(\ref{sigav}) vanishes, since the coarse-grained fields, and their power spectra, vanish for momenta larger than $1/L\to 0$, therefore only the contribution from the source remains. From eqs.~(\ref{newsig}) and (\ref{sigmaT}) we have
\beq
J_\sigma(\bk)=2\, i \, \tilde{\cal W}(\bk L)\; \int d^3 q_1 d^3 q_2\, \delta_D(\bk-\bq_1-\bq_2)\, q_2^j \,\delta v^j(\bq_1) \,\delta\phi(\bq_2)\,,
\label{Jsig}
\eeq
where $\tilde{\cal W}(\bk L)$ is the Fourier transform of the filter function.

We can approximate  the source (\ref{Jsig}) by using linear PT, which will turn out to be a good approximation in the large $L$ limit (see below). The velocity fluctuation can then be expressed in terms of its divergence as $\delta v^j(\bq) = -i q^j/q^2 \;\delta\theta(\bq)$. As for the fluctuation of the potential, we will use the Poisson equation which reads
\beq
\delta\phi(\bq)= -\frac{3}{2} \frac{{\cal H}^2}{q^2}\Omega_m\, \delta(\bq)\big(1-\tilde{\cal W}(\bq L)\big)\,.
\label{deltaPoiss}
\eeq
Putting all together, and using the property $\tilde{\cal W}(\bk L)\to 1$ as $k\to 0$, we obtain the zero-mode of the source term as
\beq
j^\sigma_0(\lambda) = - 3 {\cal H}^3\Omega_m \int d^3 q\frac{P_{\delta\,\delta\theta}(q)}{q^2}\,\big(1-\tilde{\cal W}(\bq L)\big),
\label{sorJ}
\eeq
with
\beq
\langle -\frac{\delta\theta(\bq_1)}{{\cal H}} \,\delta(\bq_2)\rangle \equiv \delta_D(\bq_1+\bq_2)P_{\delta\, \delta\theta}(q_1)\,.
\eeq
 The use of the linear approximation to the PS is justified by the $1/q^2$ factor, which improves the UV convergence of the integral in ~(\ref{sorJ}).
 In this approximation, and using the linear limit of the equations for $\delta$ and $\delta\theta$,  which give
\beq
\partial_\lambda P_{\delta\theta\,\delta\theta}(q) = -2\left(1+\frac{d \log {\cal H}}{d \lambda} \right)\,P_{\delta\theta \,\delta\theta}(q)+3 \,\Omega_m P_{ \delta\,\delta\theta}(q)\,,
\eeq
with
\beq
\langle \frac{\delta\theta(\bq_1)}{{\cal H}} \, \frac{\delta\theta(\bq_2)}{{\cal H}}\rangle \equiv \delta_D(\bq_1+\bq_2) P_{\delta\theta \,\delta\theta}(q_1)\,,
\eeq
eq.~(\ref{sigav}) can be integrated exactly, to give
\beqra
&& \bar\sigma_0(\lambda) = {\cal H}^2\,\int d^3 q\frac{P_{\delta\theta\, \delta\theta}(q)}{q^2} \big(1-\tilde{\cal W}(\bq L)\big) \to  \langle  \delta v^2(\bx)\rangle_{\mathrm{lin. th.}} \quad (\mathrm{for}\;\;L\to\infty)\,.\nonumber\\
&&\label{zero}
\eeqra 
The last equality in the equation above underlines the fact that the result coincides with the one obtained by taking the spatial average of the square of the fluctuations of the peculiar velocity, as given by linear perturbation theory. In other terms, this solution coincides, as it should, with the result which can be derived from the definition of the macroscopic velocity dispersion in eq.~(\ref{sigbar}), assuming no shell crossing has taken place ($\sigma_K^{ii}=0$).

We stress that the above computation is a non-trivial consistency check of our formalism, which is not passed by the usual way of formulating eulerian perturbation theory. Indeed, the latter implicitly assumes a vanishing coarse-graining scale, $L\to 0$. In this case, the 
zero mode of the velocity dispersion $\sigma_0$ is initially zero, and it is sourced only by the non-perturbative process of small-scale shell crossing. This is, however, inconsistent with a direct computation of $\langle \delta v^2\rangle$, which gives the non-zero result  of eq.~(\ref{zero}) already at the linear level. 
In our approach, by contrast, by changing the averaging scale $L$ we can highlight the different contributions to the velocity dispersion, namely the microscopic one and the one originated by coarse-graining. 

From the above example it is also manifest how the failure of ``traditional''  eulerian PT to reproduce this result is to be ascribed to the fact that it neglects the source terms of eqs.~(\ref{thetadot}), (\ref{sigmaT}), and (\ref{Sigma}), which carry crucial information on the short scale physics. 

Finally, we stress that this dynamical mechanism to generate a non-vanishing velocity dispersion can be entirely treated in linear PT, and is therefore completely different in nature from the non-perturbative mechanism advocated in ref.~\cite{McDonald:2009hs} to amplify an initially small $\sigma^{ij}$.

\section{Compact form} 
\label{compact}
We introduce the four-component field $\bar\vp_a(\bk,\eta)$ as
\beq
\bar\vp_a(\bk,\eta) = e^{-\eta} \left(\begin{array}{c} \bar\delta \\
-\frac{\bar\theta}{{\cal H} f}\\
\frac{k^2}{{\cal H}^2 f^2} \,\bar\sigma\\
\frac{k^2}{{\cal H}^2f^2 }\, \bar\Sigma
\end{array}\right)\,,
\label{fielddef}
\eeq 
where 
\beq
\eta\equiv \log( D_+(\tau)/D_+(\tau_{in}))\,,
\eeq and  $f=d \log D_+/d \log a$, with $D_+$ the linear growth factor. Using the equation for $f$,
\beq
\frac{d \log  {\cal H} f}{d \tau}= {\cal H}\left( \frac{3}{2}\frac{\Omega_m}{f}-f-1\right)\,,
\eeq
eqs.~(\ref{deltadot}), (\ref{thetadot}), (\ref{sigmaT}), and (\ref{Sigma}) can be cast in a compact form:
\beqra
&&\left(\delta_{ab}\partial_\eta + \Omega_{ab} \right) \bar\vp^b(\bk,\eta) =\nonumber\\
&& e^{\eta} \int d^3 q_1 d^3 q_2 \delta_D(\bk-\bq_1-\bq_2) \gamma_{abc} (k,q_1,q_2)\bar\vp_b(\bq_1,\eta)\bar\vp_c(\bq_2,\eta) - h_a(\bk,\eta)\,,\nonumber\\
&&
\label{compeq}
\eeqra
where
\beq
\Omega_{ab}(\eta)= \left(
\begin{array}{cccc}
1 & -1 & 0 & 0\\
-\frac{3}{2}\frac{\Omega_m}{f^2} &\frac{3}{2}\frac{\Omega_m}{f^2} &0 & 1\\
0 &0& 3\frac{\Omega_m}{f^2}-1&0\\
0 & 0&0& 3\frac{\Omega_m}{f^2}-1
\end{array}
\right)\,,
\label{bigomega}
\eeq
The explicit expressions for the non-vanishing components of the vertex functions are reported in Appendix A.

The four-component source term is given by
\beq
h_a(\bk,\eta) =\frac{ e^{-\eta}}{{\cal H} f} \left(\begin{array}{c} 0 \\
-\frac{1}{{\cal H}f}\,J_\theta\\
\frac{k^2}{{\cal H}^2 f^2} \, J_\sigma\\
\frac{k^2}{{\cal H}^2f^2 }\, J_\Sigma
\end{array}\right)\,.
\label{sorc}
\eeq 
The different components of the source $h_a(\bk,\eta)$ in (\ref{sorc}) can be read from  eqs.~({\ref{neweul}, \ref{newsig}) via eqs.~({\ref{thetadot}, \ref{sigmadot}), and using the Poisson equation (\ref{deltaPoiss}) to express the fluctuation of the potential, $\delta \phi$, in terms of the density field. Their explicit expressions can be found in Appendix B.

\section{Perturbation theory with sources}
\label{pertsol}

The equation of motion for the coarse-grained variables, eq.~(\ref{compeq}), can be formally solved as
\beqra
&&\bar\vp_a(\bk,\eta) = g_{ab}(\eta) \bar\vp_b^{in}(\bk) - \int_0^\eta ds\, g_{ab}(\eta-s)\, \Big[ h_b(\bk,s)  \nonumber\\
&& -  \,e^{s} \int d^3 q_1 d^3 q_2 \,\delta_D(\bk-\bq_1-\bq_2) \gamma_{bcd} (k,q_1,q_2)\bar\vp_c(\bq_1,s)\bar\vp_d(\bq_2,s)\Big]\,,\nonumber \\
&&
\label{formsol}
\eeqra
where the {\em linear propagator}
$g_{ab}(\eta-s)$ is defined as the solution of the equation \cite{RPTb}
\beq
\left(\delta_{ab}\partial_\eta + \Omega_{ab} \right) g_{bc}(\eta-s) = \delta_{ac}\,\delta_D(\eta-s)\,,
\eeq
with the retarded initial conditions, $g_{ab}(x)\to \delta_{ab}$ for $x\to 0^+$, and $g_{ab}(x)=0$ for $x < 0$.
The explicit expression of the linear propagator can be computed following the lines of Appendix A of ref.~\cite{Pietroni08}. Taking the approximation $\Omega_m/f^2\simeq 1$, one finds the analytic solution
\beq
g_{ab}(\eta)= \left({\bf B} + e^{-5\eta/2}\,{\bf A} + e^{-2\eta}\, {\bf C} \right)_{ab}\,,
\label{linprop}
\eeq
where the three matrices ${\bf A}$, ${\bf B}$, and ${\bf C}$ are given by
\beqra
&&{\bf A} = \frac{1}{5}\left(
\begin{array}{rrrr}
2 & -2 & 0 & -4\\
-3 & 3 & 0 & 6\\
0 & 0 & 0 & 0\\
0 & 0 & 0 & 0\\
\end{array}
\right)\,,\;\;\;{\bf B} = \frac{1}{5}\left(
\begin{array}{rrrr}
3 & 2 & 0 & -1\\
3 & 2 & 0 & -1\\
0 & 0 & 0 & 0\\
0 & 0 & 0 & 0\\
\end{array}
\right)\,,\nonumber\\
&&
{\bf C} = \left(
\begin{array}{rrrr}
0 & 0 & 0 & 1\\
0 & 0 & 0 & -1\\
0 & 0 & 1 & 0\\
0 & 0 & 0 & 1\\
\end{array}
\right)\,.
\eeqra
The first two terms in eq.~(\ref{linprop}) correspond, respectively, to the usual growing and decaying modes of the two-component system of density and velocity perturbations. These are excited by taking the initial perturbation field, $ \bar\vp_a^{in}(\bk)$ proportional to the vectors
\beq
u_a=\left( \begin{array}{r}
1\\1\\0\\0\end{array}
\right)\,,\qquad
v_a=\left( \begin{array}{r}
1\\-3/2\\0\\0\end{array}
\right)\,,
\label{ua}
\eeq
respectively. The third term in eq.~(\ref{linprop}) corresponds to two degenerate modes, both decaying as $\exp(-2\eta)$, which are excited by initial configurations proportional to any linear combination of the vectors
\beq
w^1_a=\left( \begin{array}{r}
0\\0\\1\\0\end{array}
\right)\,,\qquad
w^2_a=\left( \begin{array}{r}
1\\-1\\0\\1\end{array}
\right)\,.
\eeq
Notice that if the field is initially in the growing mode, $\bar\vp_a^{in}(\bk)=u_a f(\bk)$, then the evolved field, $g_{ab}(\eta) \bar\vp_b^{in}(\bk) = u_a f(\bk)$, is still in the linear growing mode even if $\Omega_m/f^2\neq 1$ in (\ref{bigomega}). 

The formal solution (\ref{formsol}) can be expanded perturbatively in powers of the vertex function $\gamma_{abc}$, 
\beq
\bar\vp_a(\bk,\eta)=\sum_{n=0}^\infty\bar\vp_a^{(n)}(\bk,\eta)\,.
\eeq
At zeroth order, we just get the first line of eq.~(\ref{formsol})
\beq
\bar\vp_a^{(0)}(\bk,\eta) = g_{ab}(\eta) \bar\vp_b^{in}(\bk) - \int_0^\eta ds\, g_{ab}(\eta-s)\, h_b(\bk,s)\,.
\label{phibar0}
\eeq
Inserting this at the r.h.s. of eq.~(\ref{formsol}) we get
\beqra
&&\bar\vp_a^{(1)}(\bk,\eta) = \int d^3q_1d^3q_2\,\delta_D(\bk-\bq_1-\bq_2) \int_0^\eta ds\, g_{ab}(\eta-s)\,e^s\,\gamma_{bcd}(k,q_1,q_2) \times \nonumber \\  
&&\qquad\qquad \bar\vp_c^{(0)}(\bq_1,s)\bar\vp_d^{(0)}(\bq_2,s)\,,
\eeqra
and, iterating the procedure, we obtain
\beqra
&&\bar\vp_a^{(2)}(\bk,\eta) = \int d^3q_1d^3q_2\,\delta_D(\bk-\bq_1-\bq_2) \int_0^\eta ds\, g_{ab}(\eta-s)\,e^s\,\gamma_{bcd}(k,q_1,q_2)\times\nonumber\\
&&\qquad\qquad\left(\bar\vp_c^{(1)}(\bq_1,s)\bar\vp_d^{(0)}(\bq_2,s)+\bar\vp_c^{(0)}(\bq_1,s)\bar\vp_d^{(1)}(\bq_2,s)\right)\,,
\eeqra 
and so on.

In the following, we will be interested in computing correlators of the perturbation fields, such as the PS
\beq
\langle \bar\vp_a(\bk_1,\eta)\bar\vp_b(\bk_2,\eta) \rangle \equiv \delta_D(\bk_1+\bk_2) \bar P_{ab}(k_1;\eta)\,,
\label{Pab}
\eeq
or the bispectrum (BS)
\beq
\langle \bar\vp_a(\bk_1,\eta)\bar\vp_b(\bk_2,\eta) \bar\vp_c(\bk_3,\eta) \rangle \equiv \delta_D(\bk_1+\bk_2+\bk_3) \bar B_{abc}(k_1,k_2,k_3;\eta)\,.
\eeq
By using the perturbative expansions above, the computation of these objects reduces to the computation of correlators between zeroth order fields, of the form
\beq
\langle \bar\vp_a^{(0)}(\bk_1,s_1)\bar\vp_a^{(0)}(\bk_2,s_2)\cdots\bar\vp_a^{(0)}(\bk_m,s_m)  \rangle\,.
\eeq
In absence of the sources $h_a$ such correlators could be decomposed in products of power spectra, under the assumption that the initial field $\bar\vp^{in}_a(\bk)$ is gaussian. If the source is non-vanishing, since it is generally non-gaussian, this is not possible any more. The only guidance to the reduction of the m-point correlators above is, besides the gaussianity of  $\bar\vp^{in}_a(\bk)$, the vanishing of the source and field zero modes (notice that the $h_3$ and $h_4$ components contain a factor $k^2$ with respect to $J_\sigma$ and $J_\Sigma$, see (\ref{sorc}))
\beq
\langle \bar\vp_a^{(0)}(\bk,\eta)\rangle= \langle \bar\vp^{in}_a(\bk) \rangle= \langle h_a(\bk,s) \rangle = 0\,.
\eeq
In order to compute the PS to $O(\gamma^2)$ we need to take into account the {\em generalized} PS, BS, and trispectrum (TS), defined respectively as
\beqra
&&\langle \bar\vp_a^{(0)}(\bk_1,s_1)\bar\vp_b^{(0)}(\bk_2,s_2) \rangle \equiv \delta_D(\bk_1+\bk_2) {\cal P}^0_{ab}(k_1;s_1,s_2)\,,\nonumber\\
&&\langle \bar\vp_a^{(0)}(\bk_1,s_1)\bar\vp_b^{(0)}(\bk_2,s_2)\bar\vp_c^{(0)}(\bk_3,s_3) \rangle \equiv \nonumber\\
&&\qquad\qquad\qquad\qquad \delta_D(\bk_1+\bk_2+\bk_3) {\cal B}^0_{abc}(k_1,k_2,k_3;s_1,s_2,s_3)\,,\nonumber\\
&& \langle \bar\vp_a^{(0)}(\bk_1,s_1)\bar\vp_b^{(0)}(\bk_2,s_2)\bar\vp_c^{(0)}(\bk_3,s_3) )\bar\vp_d^{(0)}(\bk_4,s_4) \rangle_{\mathrm{conn.}} \equiv\nonumber\\
&&\qquad\qquad\qquad\qquad \delta_D(\bk_1+\bk_2+\bk_3+\bk_4) {\cal T}^0_{abcd}(\bk_1,\bk_2,\bk_3,\bk_4;s_1,s_2,s_3,s_4)\,,\nonumber
\\&& \label{corres}
\eeqra
where the trispectrum, as usual, is defined as the {\it connected} part of the four-point correlator.
In terms of these quantities, the PS (\ref{Pab}) is given by
\beqra
 &&\bar P_{ab}(k;\eta)= {\cal P}^0_{ab}(k;\eta,\eta)\nonumber\\
 &&+ \int d^3q_1d^3q_2\,\delta_D(\bk-\bq_1-\bq_2) \int_0^\eta ds \,g_{ac}(\eta-s) \nonumber\\
 &&\times \Bigg\{\bigg[e^s \gamma_{cde}(k,q_1,q_2)  {\cal B}^0_{bde}(k,q_1,q_2;\eta,s,s) + (a\leftrightarrow b)\bigg]\nonumber\\
 &&\qquad + 2 \int_0^\eta ds^\prime  g_{bd}(\eta-s^\prime) e^{s+s^\prime}\times\nonumber\\
 &&\qquad\qquad\gamma_{cef}(k,q_1,q_2)\gamma_{dgh}(k,q_1,q_2) {\cal P}^0_{eg}(q_1;s,s^\prime){\cal P}^0_{fh}(q_2;s,s^\prime)\nonumber\\
 && \qquad+ \bigg[4 \int_0^s ds^\prime g_{df}(s-s^\prime)e^{s+s^\prime}\times\nonumber\\
  &&\qquad\qquad\gamma_{cde}(k,q_1,q_2)\gamma_{fgh}(q_1,q_2,k) {\cal P}^0_{eg}(q_2;s,s^\prime){\cal P}^0_{hb}(k;s^\prime,\eta) 
  + (a\leftrightarrow b)  \bigg] \Bigg\}\nonumber\\
  &&+ \int d^3q_1d^3q_2d^3p_1d^3p_2\, \delta_D(\bk-\bq_1-\bq_2)  \int_0^\eta ds \,g_{ac}(\eta-s) \nonumber\\
  &&\qquad\times \Bigg\{ \int_0^\eta ds^\prime g_{bd}(\eta-s^\prime) e^{s+s^\prime} \delta_D(\bk+\bp_1+\bp_2)\times\nonumber\\
  &&\qquad\qquad\gamma_{cef}(k,q_1,q_2)\gamma_{dgh}(k,p_1,p_2) {\cal T}^0_{efgh}(q_1,q_2,p_1,p_2;s,s,s^\prime,s^\prime)\nonumber\\
 && \qquad\qquad+\bigg[\int_0^s ds^\prime g_{df}(s-s^\prime)e^{s+s^\prime} \delta_D(-\bq_1+\bp_1+\bp_2)\times\nonumber\\
  &&\qquad\qquad\gamma_{cde}(k,q_1,q_2)\gamma_{fgh}(q_1,p_1,p_2) {\cal T}^0_{gheb}(p_1,p_2,q_2,k;s^\prime,s^\prime,s,\eta) + (a\leftrightarrow b)\bigg]\,,\nonumber\\
\label{psbar}
\eeqra
up to $O(\gamma^2)$ terms.

Using eq.~(\ref{phibar0}) in (\ref{corres}), we realize that in order to compute the PS in (\ref{psbar}) we need to evaluate correlators such as
\beq
\langle \bar\vp^{in}_a(\bk)  \bar\vp^{in}_b(\bk^\prime) \rangle\,,\qquad \langle \bar\vp^{in}_a(\bk) h_b(\bk^\prime,s) \rangle\,,\qquad \langle h_a(\bk,s) h_b(\bk^\prime,s^\prime) \rangle\,,
\eeq 
as well as correlators involving more than three $\bar\vp^{in}$ fields and $h$ sources. 
To compute the correlators involving only $\bar\vp^{in}$ fields we assume that the initial conditions are taken at a redshift high enough that the scales larger than the coarse-graining one, {\it i.e.} such that $k \alt 1/L$, are well in the linear regime, and perturbations are in the growing mode identified by the $u_a$ vector in eq.~(\ref{ua}). Therefore, we have
\beq
\langle \bar\vp^{in}_a(\bk)  \bar\vp^{in}_b(\bk^\prime) \rangle = \delta_D(\bk+\bk^\prime) \bar P_{ab}^{in} (k) =  \delta_D(\bk+\bk^\prime) \tilde{W}^2(\bk L)  P^{0}(k) u_a u_b\,,
\label{psin}
\eeq
where $P^0(k)$ is the linear density-density PS evaluated at the initial redshift. Notice the $\tilde{W}^2(\bk L)$ factor, making the only difference between the linear PS's for the averaged quantities and those for the not-averaged ones. At $z=z_{in}$ we neglect the quadratic contribution $O(\delta n/\bar{n} \; v^i)$ in the integral in (\ref{vbari}) with respect to the linear one. As usual, if the initial field is non-gaussian, also the initial bispectra, trispectra, and so on should be given, otherwise, eq.~(\ref{psin}) completely defines its statistics. In this paper we will work under this assumption, leaving the discussion of primordial non-gaussianities to the future.

The contribution to the $O(\gamma^2)$ PS  from correlators involving only $\bar\vp^{in}$ fields ({\it i.e.} for $h_a\to0$) can be obtained by setting ${\cal P}^0_{ab}(\bq;s,s^\prime) \to \tilde{W}^2(\bq L) P^0(q) u_a u_b$ and $ {\cal B}^0_{abc}=  {\cal T}^0_{abcd}=0$  in eq.~(\ref{psbar}). This gives the usual expression for the 1-loop PS (see for instance eq.~(162) of ref.\cite{PT}), in which the linear PS have been multiplied by the UV filtering functions $\tilde{W}^2$. Therefore, as expected, the contribution from small scale fluctuations are entirely encoded in the sources. In the next section, we will show how to compute the sources correlators  perturbatively in order to reproduce the full 1-loop result, while, in Section~\ref{sournbody}, we will compute the PS components from a N-body simulation, and compare with our results.

\section{Perturbative expansion of sources}
\label{pertsourc}

In this section, we will compute the PS for the coarse-grained field, $\bar P_{ab}$, up to second order, {\it i.e.} keeping contributions up to $O((\bar P^{in})^2,\, \bar P^{in}\delta P^{in},\, (\delta P^{in})^2)$, where $ \bar P^{in}(k) =  \tilde{W}^2(\bk L)\, P^{in}(k) $ and 
$ \delta P^{in}(k) =  (1- \tilde{W}(\bk L))^2\, P^{in}(k) $, using the explicit expression for the sources $h_a(\bk,\eta)$ given in Appendix B. 

In order to simplify the computation, we will present explicit formulae for the sharp cut-off limit, namely, we will consider the filter function 
\beq
\tilde{W}_\mathrm{sharp}(k L) = \Theta\left(1 - k L/(2 \pi) \right)\,,
\eeq
for which the following property holds,
\beq
 \tilde{W}_\mathrm{sharp}(\bk L) \left(1-\tilde{W}_\mathrm{sharp}(\bk L) \right) = 0\,.
 \label{sharp}
\eeq
However, in Section~\ref{sournbody}, we will also present results obtained with a smooth cut-off.

We will compute the different components of PS for the $\vpb_a$ fields by expanding also the 
$\delta\vp_a$ fields appearing in the sources (\ref{h2}), (\ref{h34}) perturbatively. It can be done by solving the
equations of motion for $\delta\vp_a$, which can be read out from eqs.~(\ref{vlasovdeltaC}) and (\ref{deltaPoiss}),
\beqra
&&\left(\delta_{ab}\partial_\eta + \Omega_{ab} \right) \delta\vp^b(\bk,\eta) = h_a(\bk,\eta) \nonumber\\
&&\qquad+ e^{\eta} \int d^3 q_1 d^3 q_2 \delta_D(\bk-\bq_1-\bq_2) \bigg[\gamma_{abc} (k,q_1,q_2)\bigg(2\, \bar\vp_b(\bq_1,\eta)\delta\vp_c(\bq_2,\eta) \nonumber\\
&&\qquad+\delta\vp_b(\bq_1,\eta)\delta\vp_c(\bq_2,\eta) \bigg) \bigg]\,.\nonumber\\
&&
\label{compeqdelta}
\eeqra
Notice that the source term $h_a$ comes with a positive sign, differently from the case of eq.~(\ref{compeq}) for $\vpb_a$. 
In the sharp cut-off limit, the field $\vpb_a(\bk,\eta)$ vanishes for $k > 2 \pi/L$, therefore we are interested in computing the correlators at momenta $k<2 \pi/L$.

At the desired order, the terms of eq.~(\ref{psbar}) contributing to the PS are
\beqra
 &&\bar P_{ab}(k;\eta)=  u_a u_b \,\bar{P}^{in} (k)\nonumber\\
 &&- \bigg[ g_{ac}(\eta)\int_0^\eta ds\, g_{bd}(\eta-s) \,\langle \vpb^{in}_c(\bk) h_d(\bk^\prime, s)\rangle/\delta_D(\bk+\bk^\prime) + (a\leftrightarrow b)\bigg]\nonumber\\
 &&+ \int_0^\eta ds\, ds^\prime\, g_{ac}(\eta-s) g_{bd}(\eta-s^\prime) \langle h_c(\bk, s) h_d(\bk^\prime, s^\prime)\rangle/\delta_D(\bk+\bk^\prime)\nonumber\\
&& + \int d^3q_1d^3q_2\,\delta_D(\bk-\bq_1-\bq_2) \int_0^\eta ds \,g_{ac}(\eta-s) \nonumber\\
 && \times \Bigg\{ 2 \bar P^{in}(q_1) \bar P^{in}(q_2)\int_0^\eta ds^\prime  g_{bd}(\eta-s^\prime) e^{s+s^\prime}\gamma_{cef}(k,q_1,q_2)\gamma_{dgh}(k,q_1,q_2)  u_eu_g u_f u_h \nonumber\\
 && + \bigg[4 \bar P^{in}(k) \bar P^{in}(q_2)\int_0^s ds^\prime g_{df}(s-s^\prime)e^{s+s^\prime}\gamma_{cde}(k,q_1,q_2)\gamma_{fgh}(q_1,q_2,k) u_e u_g u_h u_b
  + (a\leftrightarrow b)  \bigg] \Bigg\}\,.\nonumber\\
  \label{p1l}
\eeqra
The last three lines give the usual expression for the 1-loop PS in which the initial PS, that is $P^{in}$, is replaced by the one for the coarse-grained fields, $\bar P^{in}$. Therefore, the corresponding momentum integrals are cut-off for $q \agt 2 \pi/L$. 

The small-scale contribution comes from the correlators involving the sources, keeping only terms quadratic in $ \bar P^{in}(k) $ and $ \delta P^{in}(k)$. The $O(\delta {P^{in}}^2)$ term is obtained from the $\langle h_c h_d \rangle$ correlator by using the source terms  (\ref{h2})-(\ref{h34}) with $\delta \vp_a (\bq,\eta) =  u_a \delta \vp^{in} (\bq)$. It gives exactly the same expression as the fifth line of eq.~(\ref{p1l}), in which both the coarse-grained PS are replaced with $\delta P^{in}$. 

The $O(\bar P^{in} \delta P^{in})$ contribution comes from the  $\langle \vpb_c^{in}\, h_d \rangle$ correlator, in which the $\delta \vp$ contained in the first lines of eqs.~(\ref{h2}) and (\ref{h34}) have to be computed to the first perturbative order according to eq.~(\ref{compeqdelta}). 
As a result, the expression for the 1-loop PS for the coarse-grained field is given by 
\beqra
\bar P_{11}(k,\eta)=&&  \bar P^{in}(k) \nonumber\\
&&+ 2 e^{2\eta}\bar P^{in}(k)  \left[\Delta G[\bar P^{in};k]+  \tilde{W}(\bk L)\left(\Delta G[\delta P^{in};k] +\frac{1}{6}k^2 \,\delta \sigma_v^2 \right)\right] \nonumber\\
&&+e^{2\eta} \left(\Delta P^{MC}[\bar P^{in};k]+ \tilde{W}(\bk L)^2 \Delta P^{MC}[\delta P^{in};k]\right)\,,
\label{p1loop}
\eeqra
for the $1-1$ ({\it i.e.} density-density) component, where
\beqra
&&\Delta G[ P;k] \equiv  \frac{k^3\pi}{252}\int_0^\infty dr\, P(k r)  \nonumber\\
&&\quad\quad\times\bigg(\frac{12}{r^2} - 158+100 r^2 -42 r^4 +\frac{3}{r^3}(r^2-1)^3(2+7r^2) \log\left|\frac{1+r}{1-r}\right|\bigg),\nonumber\\
&&
\eeqra
\beq
\delta \sigma_v^2 \equiv \frac{1}{3} \int d^3q \frac{\delta P^{in}(q)}{q^2}\,,
\eeq
and 
\beqra
\Delta P^{MC}[P;k] =&& \frac{k^3\pi }{49}\int_0^\infty dr P(kr)\int_{-1}^{1} dx P(k\sqrt{1+r^2-2 r x})\nonumber\\
&& \qquad\quad\times\frac{(3r+7x -10 r x^2)^2}{(1+r^2-2 r x)^2}\,.
\eeqra
The $\delta P^{in}$ dependent terms in (\ref{p1loop}) come from the contributions to eq.~(\ref{p1l}) containing the sources.
Since we are working in the sharp cut-off limit, see eq.~(\ref{sharp}), the sum of the momentum integrals in the  two $\Delta G$-terms and in the two $\Delta P^{MC}$ ones in (\ref{p1loop}) reproduces the standard one-loop result for the PS at momenta $k<2 \pi/L$. The only difference then comes from the $1/6\, k^2 \delta\sigma_v^2$ term, which vanishes in the limit of vanishing coarse-graining scale, $L\to 0$, as $\delta P^{in}\to 0$ (and $\bar P^{in}\to P^{in}$) in that limit.

A non-vanishing $L$ gives rise to non-vanishing velocity dispersion. Indeed, the cross-correlator between $\vp_1$ and $\vp_3$ is given by
\beqra
\bar P_{13}(k,\eta)=
&& e^{2\eta}\left(\bar P^{in}(k)  \tilde{W}(\bk L)\Delta G_{13}[\delta P^{in};k] + \tilde{W}^2(\bk L) \Delta P^{MC}_{13}[\delta P^{in};k]\right)\,,\nonumber\\
&&
\label{p1loop13}
\eeqra
with
\beqra
&&\Delta G_{13}[ P;k] \equiv - \frac{k^3\pi}{7}\int_0^\infty dr\, P(k r)  \nonumber\\
&&\quad\quad\times\bigg[\frac{6}{r^2}-30-6 r^2+\frac{3}{r^3}(r^2-1)^3 (1+r^2) \log\left|\frac{1+r}{1-r}\right|\bigg],\nonumber\\
&&
\eeqra
and 
\beqra
\Delta P^{MC}_{13}[P;k] =&&- \frac{2 k^3\pi }{7}\int_0^\infty dr P(kr)\int_{-1}^{1} dx P(k\sqrt{1+r^2-2 r x})\nonumber\\
&& \qquad\quad\times\frac{(3r+7x -10 r x^2)(x-r)}{(1+r^2-2 r x)^2}\,, 
\eeqra
and similarly for the $\vp_1-\vp_4$ cross-correlator, 
\beqra
\bar P_{14}(k,\eta)=
&& e^{2\eta}\left(\bar P^{in}(k)  \tilde{W}(\bk L)\Delta G_{14}[\delta P^{in};k] + \tilde{W}^2(\bk L) \Delta P^{MC}_{14}[\delta P^{in};k]\right)\,,\nonumber\\
&&
\label{p1loop14}
\eeqra
with
\beqra
&&\Delta G_{14}[ P;k] \equiv - \frac{k^3\pi}{14}\int_0^\infty dr\, P(k r)  \nonumber\\
&&\quad\quad\times\bigg[\frac{6}{r^2}-38+10 r^2-6r^4+\frac{3}{r^3}(r^2-1)^3 (1+r^2) \log\left|\frac{1+r}{1-r}\right|\bigg],\nonumber\\
&&
\eeqra
and 
\beqra
\Delta P^{MC}_{14}[P;k] =&&\frac{2 k^3\pi }{7}\int_0^\infty dr P(kr)\int_{-1}^{1} dx P(k\sqrt{1+r^2-2 r x})\nonumber\\
&& \qquad\quad\times\frac{(3r+7x -10 r x^2)(r x-1) x}{(1+r^2-2 r x)^2}\,.
\eeqra
\begin{figure}
\centerline{\includegraphics[width = 16cm,keepaspectratio=true]{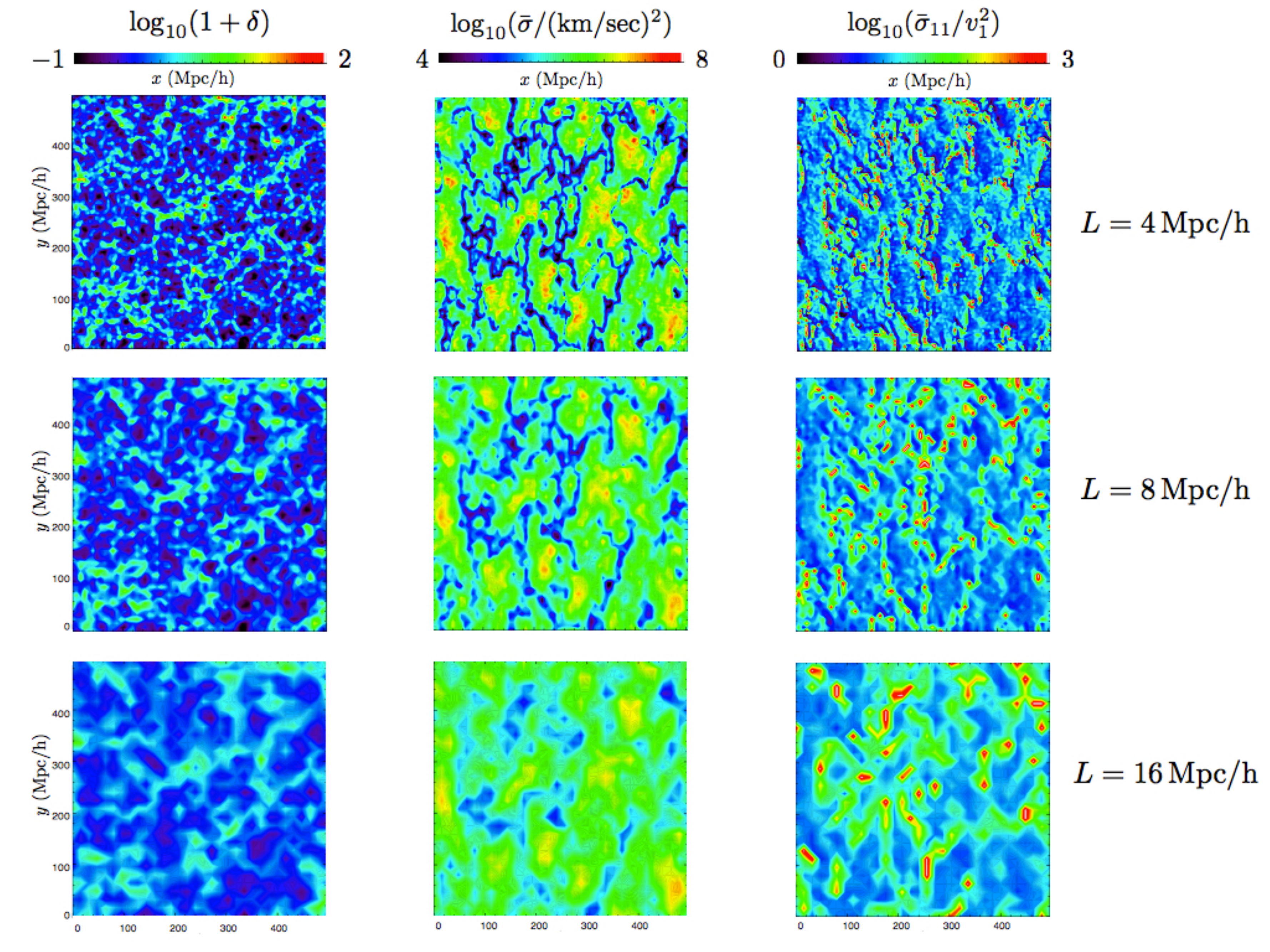}}
\caption{Snapshots of various quantities obtained from the simulation described in the text. The left column is the density field, the central one is the trace of the velocity dispersion tensor, and the right one the ratio $\bar{\sigma}_{11}/\bar{v}_1^2$. The different lines are obtained by taking grids of spacing $L=4,\,8,$ and $16\,\mathrm{Mpc}/h$, from top to bottom.}
\label{3Slices}
\end{figure}

\section{Comparison with simulations}
\label{sournbody}
In order to compare our calculations with the numerical results of
N-body simulations we perform a cosmological simulation using the
public available Tree-Particle Mesh {\small{GADGET-2}} code
\cite{volker05}. The simulation consists of a periodic volume of
linear size 512 $h^{-1}$ comoving Mpc which is evolved from $z=90$ to
$z=0$, this volume is filled with $512^3$ dark matter particles and no
baryons and cooling or astrophysical processes are included.  The
cosmological reference model is a $\Lambda$CDM with the following
parameters: $\Omega_{\rm 0m}=0.3$, $\Omega_{\rm 0\Lambda}=0.7$,
$h=0.7$, $n_{\rm s}=0.95$ and $\sigma_8=0.85$. The initial conditions
are generated using the Eisenstein \& Hu fitting formula
\cite{eisensteinhu}. The
particle mesh grid used to calculate the long range forces is chosen
to be $512^3$, while the gravitational softening is 60 kpc$/h$
in comoving units (this value is constant in redshift).

At each snapshot we extract the positions and peculiar velocities of the dark matter particles and we interpolate
using a Cloud-In-Cell algorithm these quantities on a $128^3$
grid. We then compute the mass density, the center of mass velocity and the velocity dispersion tensor for each elementary cube of size $(4 \,\mathrm{Mpc}/h)^3$, corresponding to the quantities given by eqs.~(\ref{nmic}), (\ref{vav}) or (\ref{vbari}), and (\ref{sigbar0}) or (\ref{sigbar}), respectively. Then we group these ``elementary" cubes in larger cubes, of size $(8 \,\mathrm{Mpc}/h)^3$ or $(16 \,\mathrm{Mpc}/h)^3$, and for each of these larger cubes we compute the mass density, the c.o.m. velocity and the velocity dispersion tensor. 

The decomposition of the velocity dispersion given in eq.~(\ref{sigbar}) illustrates a difference between the elementary cubes of of size $(4 \,\mathrm{Mpc}/h)^3$ and the larger ones. For the elementary cubes, the velocity dispersion is entirely given by the contribution of the second line, namely by the fact that the different particles inside a cube have different velocities, whereas the ``microscopic" contribution, at this level, would be given by the velocity dispersion of sub-particles inside each particle, which we do not resolve. When considering larger cubes, on the other hand, the velocity dispersion associated to each elementary cube contributes to the ``microscopic" term at the first line of  eq.~(\ref{sigbar}), whereas the dispersion of the center of mass velocities of the different elementary cubes contributes to the second line. The procedure could be iterated {\it ad-libitum}, with the total velocity dispersion at a given volume size playing the role of the microscopic velocity dispersion at larger volumes.

In Figure \ref{3Slices} we qualitatively show some physical
quantities extracted from the $z=0$ output: each line corresponds to the same slice of the $(512\, \mathrm{Mpc}/h)^3$ simulation seen at a given coarse-graining scale. The first column shows the dark matter density, the second the trace of the velocity dispersion tensor and the third shows the ratio $\bar{\sigma}^{11}/\bar{v}_1^2$.

Following the first column downwards we verify the obvious expectation that increasing the coarse-graining scale the density fluctuation field becomes less nonlinear, and therefore PT gets more and more well behaved. 

From the second column we see that velocity dispersion is ubiquitous and, as expected, correlated with matter density. However, the relevant quantities to gauge if the single stream approximation is well motivated are the ratios $\bar{\sigma}^{ij}/\bar{v}_i\bar{v}_j$, see for instance, eq.~(\ref{2mom}). From the top panel in the third column we see that, as expected, the single stream approximation badly fails inside overdense regions and, at $z=0$ it is marginally acceptable elsewhere (the  $\bar{\sigma}^{11}/\bar{v}_1^2$ is everywhere $\agt 1$). However, increasing the coarse-graining scale $L$, we see that the regions where the single stream approximation fails get enlarged, namely, the single stream approximation gets worse at larger scales. 
Therefore, the content of the snapshots in Figure \ref{3Slices} can be qualitatively summarized as follows. At small coarse-graining scale the single stream approximation is acceptable over a large portion of the volume, but the density field is highly nonlinear and therefore PT is unreliable. On the other hand, increasing the coarse-graining scale one gets a more linear density field, but velocity dispersion becomes non-negligible at larger scales and therefore the PT scheme should be enlarged in order to take it into account.

\begin{figure}
\centerline{\includegraphics[width = 16cm,keepaspectratio=true]{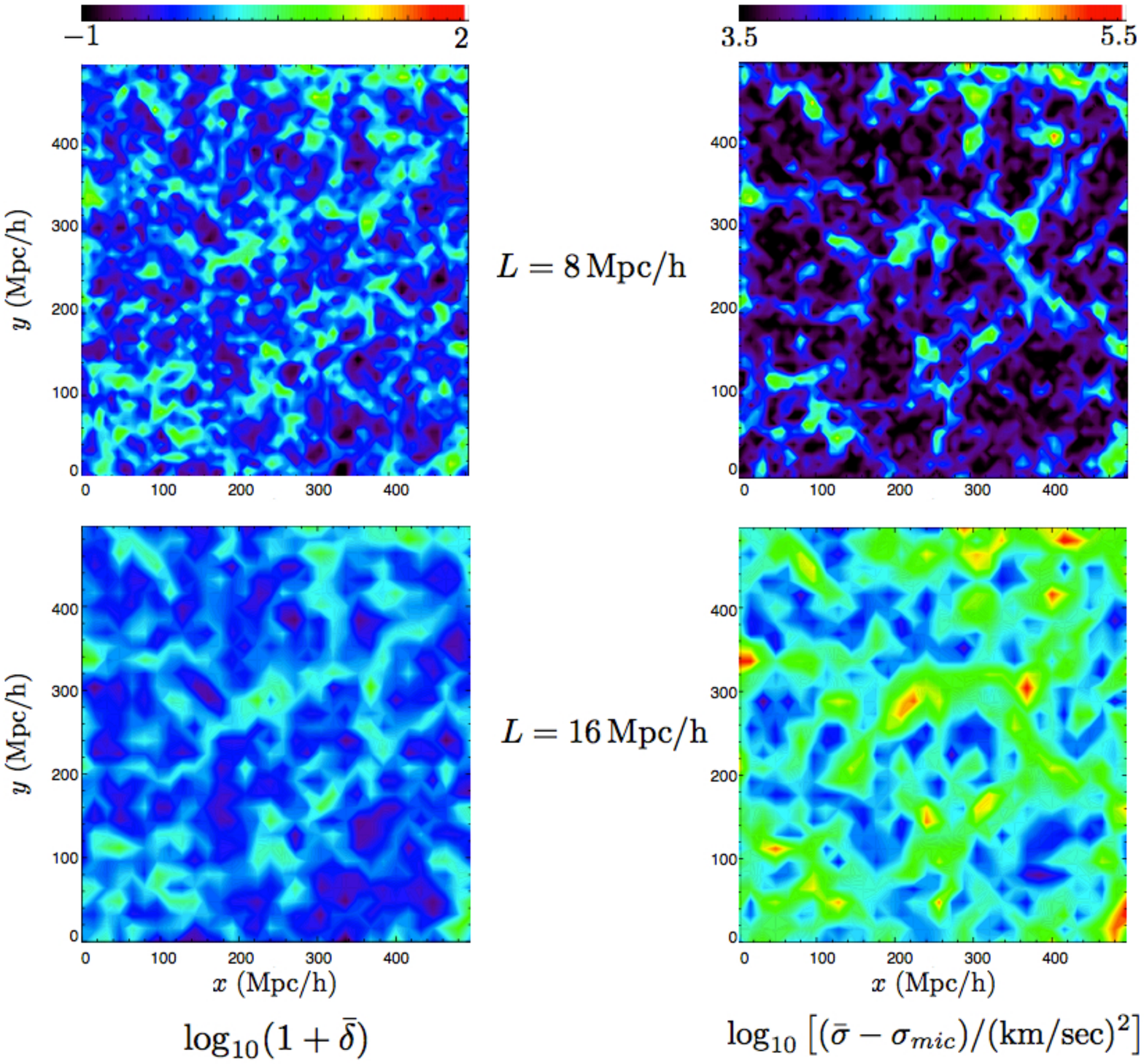}}
\caption{On the left column the same density field as in Figure \ref{3Slices} is plotted. The right column represents the macroscopic contribution to velocity dispersion, as defined by the second line of eq.~(\ref{sigbar}).}
\label{2Slices}
\end{figure}

In Figure~\ref{2Slices} we plot two quantities which can be compared to the analytical results presented in the previous section. On the left column we have again the density field, already shown in Figure~\ref{3Slices}. On the right column we have the ``macroscopic" contribution to the trace of the velocity dispersion tensor, {\it i.e.} the second line of eq.~(\ref{sigbar}). Indeed, computing the sources in PT, as we did in Section~\ref{pertsourc}, amounts to neglecting the ``microscopic" contribution to velocity dispersion (as well as to any other deviation from the single stream approximation), therefore the first line of eq.~(\ref{sigbar}) is implicitly taken to vanish in this computation. On the other hand, it could be included by measuring the sources directly from simulations and then following the general scheme described in Section~\ref{pertsol}. This procedure will be described in detail in a forthcoming paper. 

The right column in Figure~\ref{2Slices} shows the correlation between the macroscopic velocity dispersion and the density field, which is even more evident than for the total velocity dispersion. Moreover, we clearly see that increasing the coarse-graining scale $L$ the macroscopic velocity dispersion increases.
These results are reproduced by our analytical computations, as shown in Figure~\ref{Ptutti}. There, we plot the density-density PS, $P_{11}$, and the cross-correlator between the density and the trace of the macroscopic velocity dispersion, $P_{13}$, for different redshifts, $z=1.5, \,0$, and coarse-graining scales, $L=8,\,16\,\mathrm{Mpc}/h$. The dots represent the corresponding cross-correlators extracted from the simulations.

A relevant technical issue is the way one implements the coarse-graining procedure. As far as simulations are concerned, the natural way to proceed is, as described above, to take larger and larger cubes, that is, to use a cubic cut-off in real space. On the other hand, when it comes to the analytical computations, it is more practical to work in momentum space. Here there is no privileged way to take a cut-off. In our computations, we used two different schemes: a sharp cut-off,
\beq
\tilde{W}_\mathrm{sharp}(k L) = \Theta\left(1 - k L/(2 \pi) \right) \,,
 \eeq
 and a smooth one
\beq
\tilde{W}_\mathrm{smooth}(k L) = 3\left(\frac{\sin(k L/2)}{(k L/2)^3} -\frac{\cos(k L/2)}{(k L/2)^2} \right)\,,
\eeq
the latter being the Fourier transform of the spherical top-hat filter with radius $L/2$.

From Figure~\ref{Ptutti} we see that, at momentum scales approaching $2 \pi/L$, the difference between the two cut-off schemes manifests itself in our results both for $P_{11}$ and $P_{13}$. At a larger coarse-graining scale, $L=16\,\mathrm{Mpc}/h$, the results obtained with the smooth cut-off trace the $P_{11}$ from simulations  better than those obtained with the sharp cut-off, which  -- apart from the $\frac{1}{6}k^2 \,\delta \sigma_v^2$ term at the second line of eq.~(\ref{p1loop}) -- coincides with the standard 1-loop computation. The cut-off dependence is, however, strongly alleviated by taking ratios of quantities computed in the same cut-off scheme, as we show in Figure~\ref{Pratios}, where the ratio $P_{13}/P_{11}$ is plotted.

\begin{figure}
\centerline{\includegraphics[width = 8cm,keepaspectratio=true]{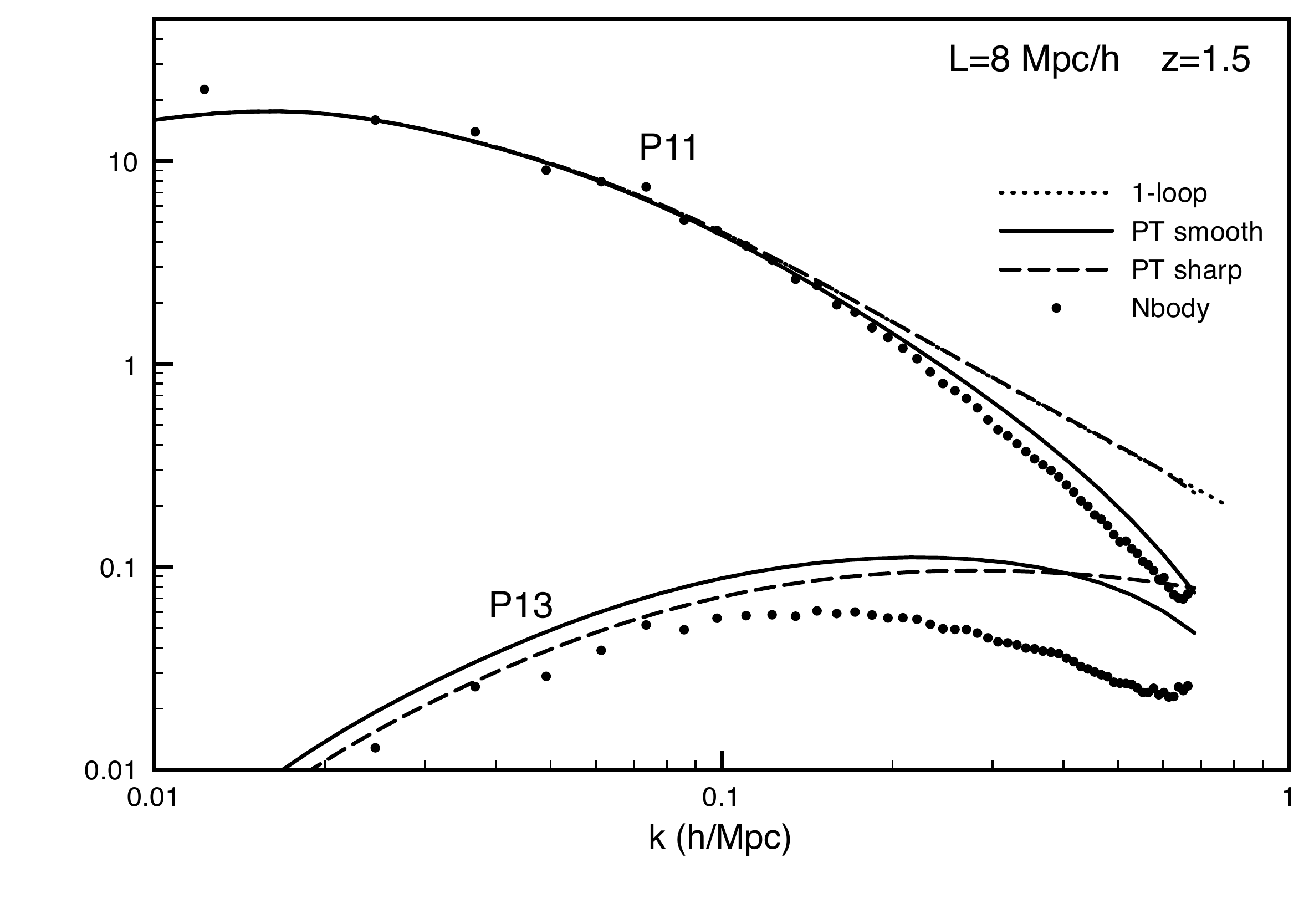}
\includegraphics[width = 8cm,keepaspectratio=true]{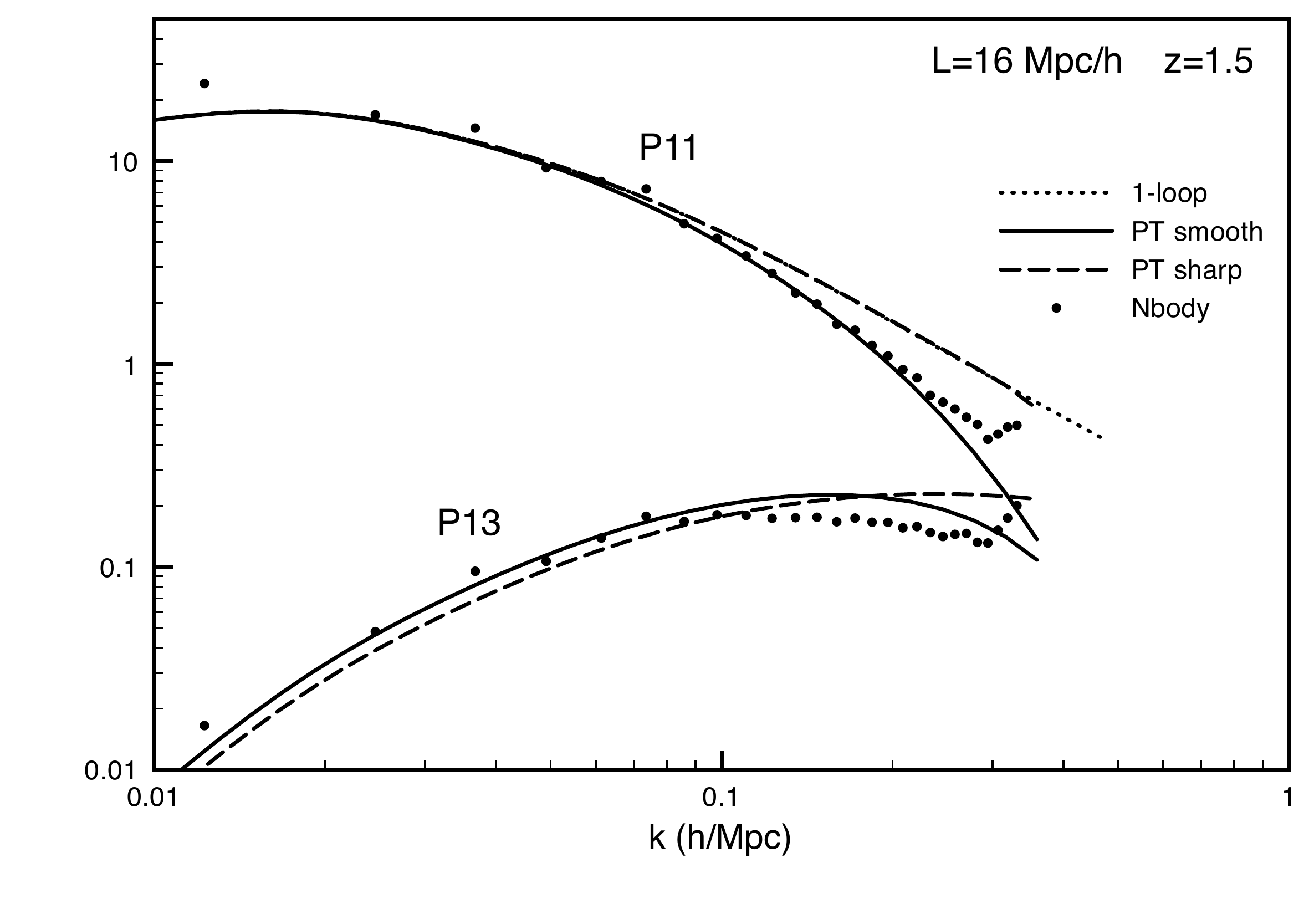}}
\centerline{\includegraphics[width = 8cm,keepaspectratio=true]{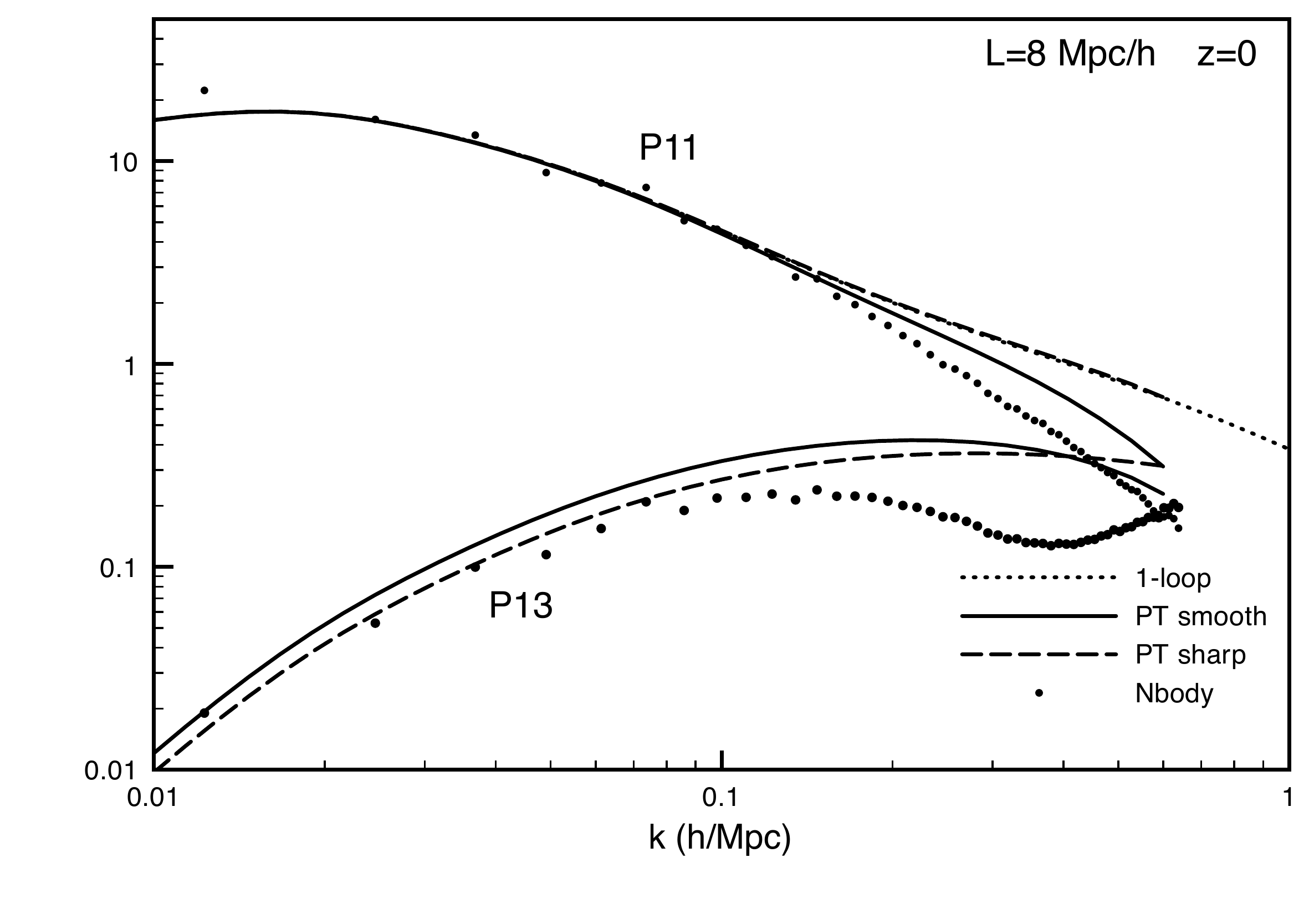}
\includegraphics[width = 8cm,keepaspectratio=true]{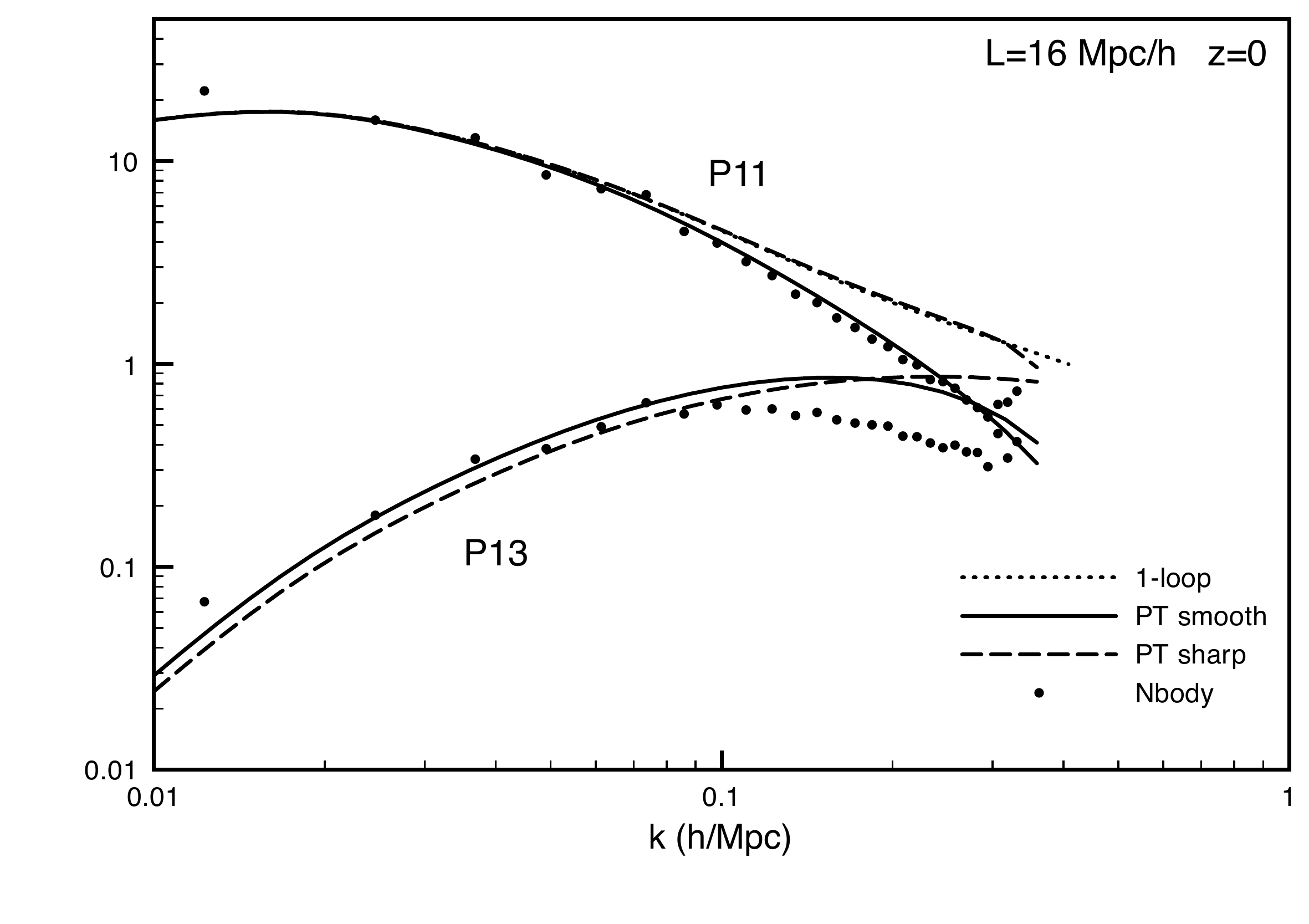}}
\caption{Comparison between the perturbative computation described in Section~\ref{pertsourc} and simulations. $P_{11}$ represents the density field PS, whereas $P_{13}$ is the cross-correlator between density and the macroscopic component of the trace of the velocity dispersion tensor. The fields have been rescaled according to eq.~(\ref{fielddef}). The continuos (dashed) lines represent the results obtained with the smooth (sharp) cut-off. The dotted line is the standard 1-loop result, while the dots are obtained from N-body simulations. }
\label{Ptutti}
\end{figure}

\begin{figure}
\centerline{\includegraphics[width = 8cm,keepaspectratio=true]{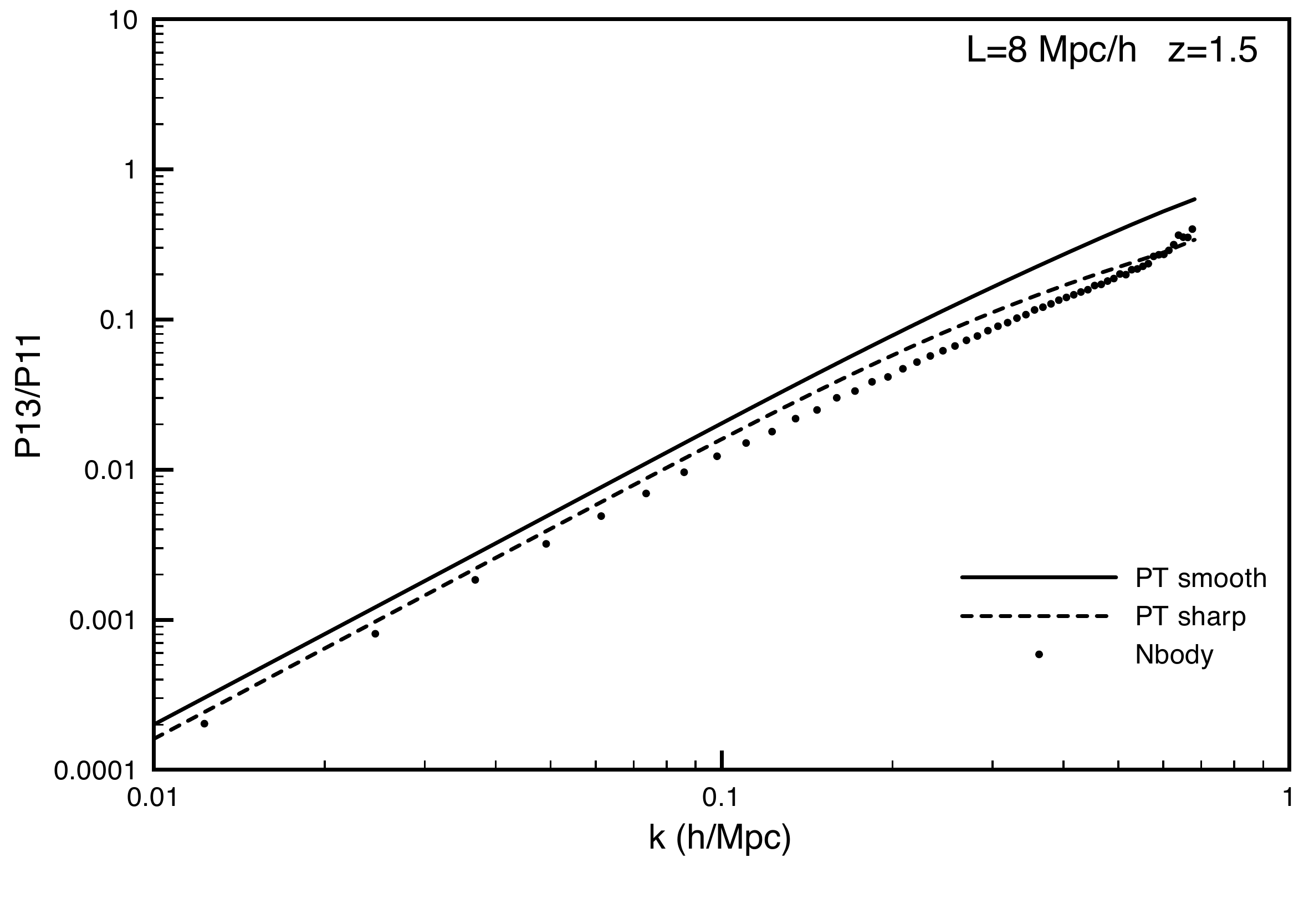}
\includegraphics[width = 8cm,keepaspectratio=true]{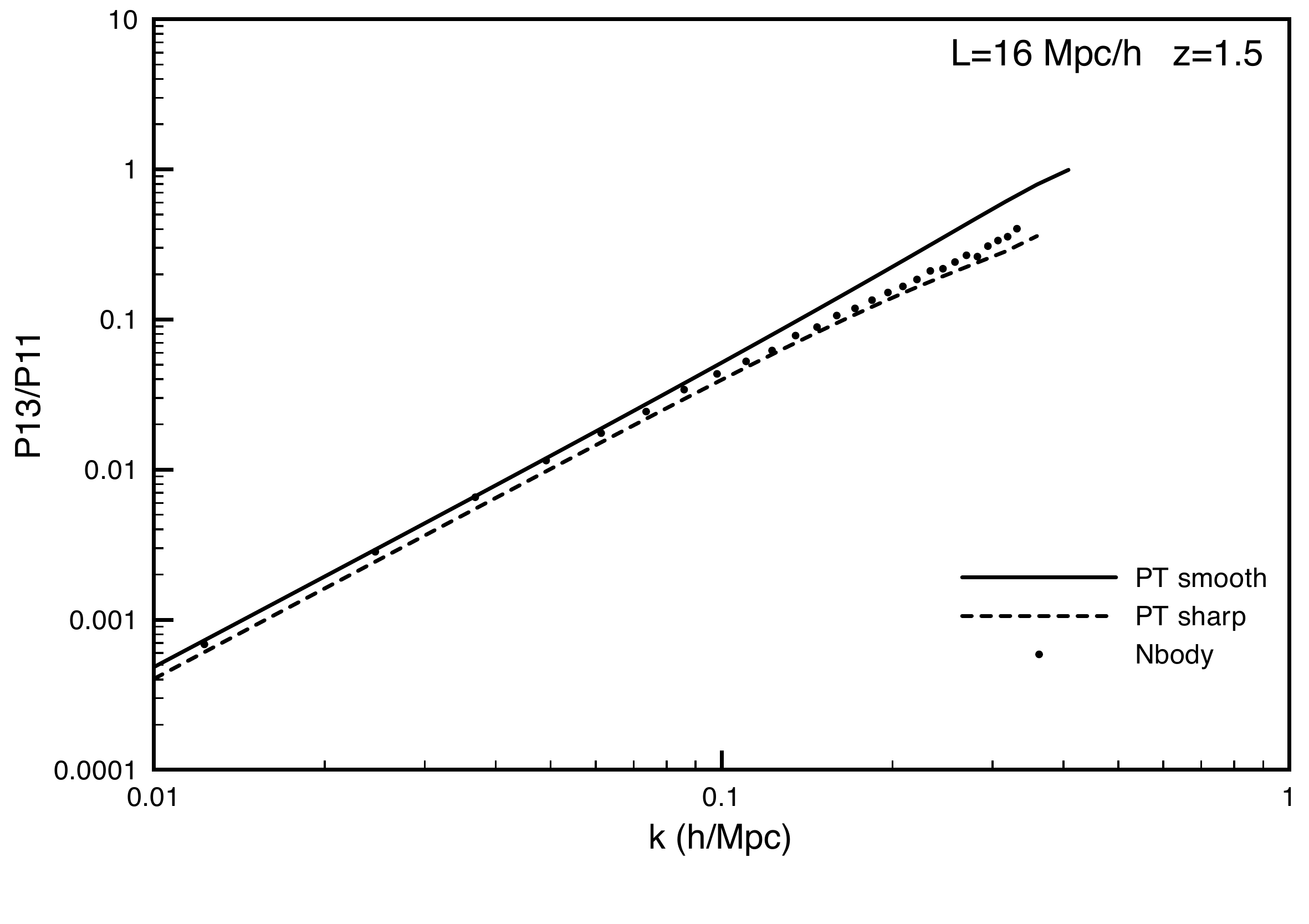}}
\centerline{\includegraphics[width = 8cm,keepaspectratio=true]{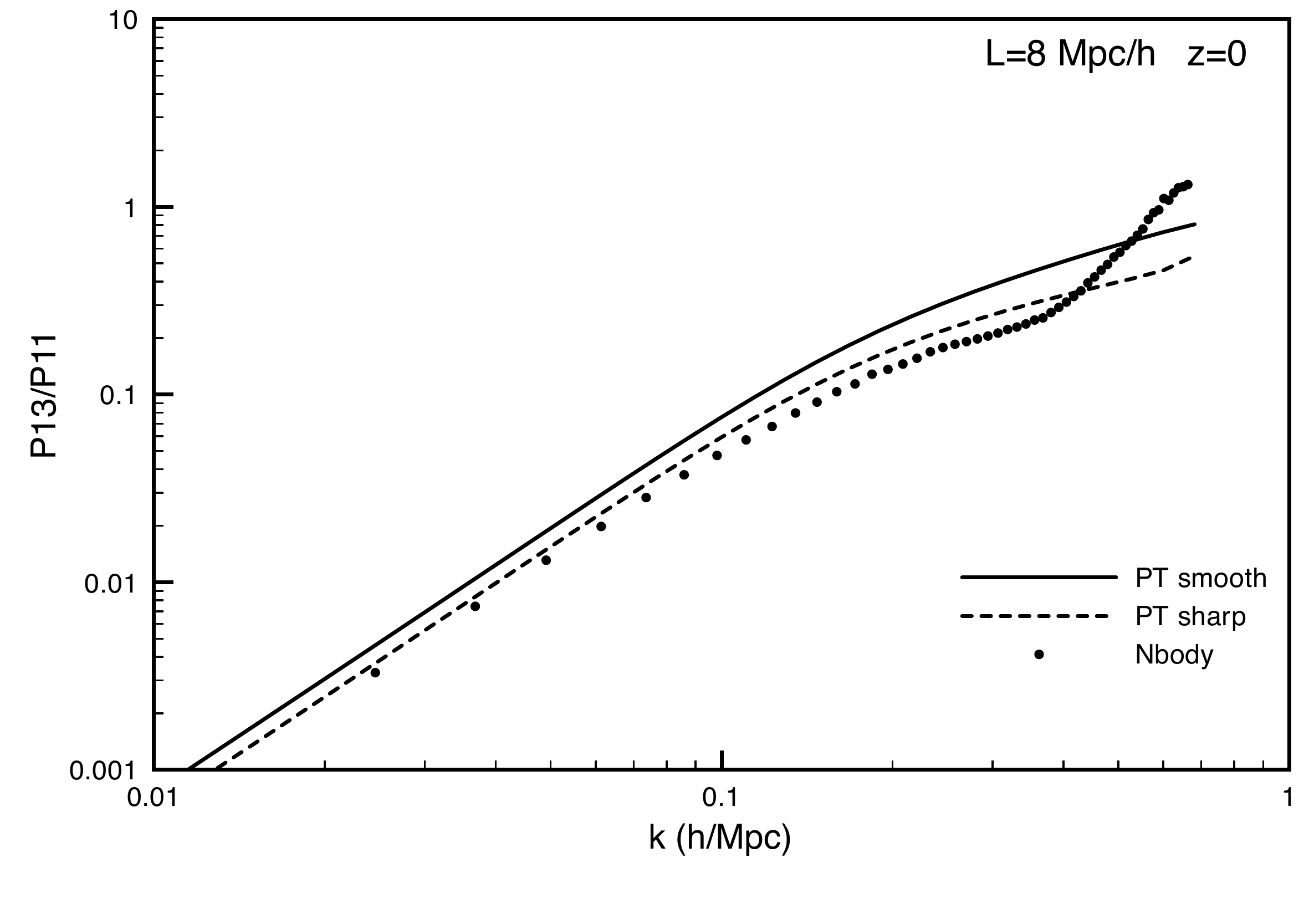}
\includegraphics[width = 8cm,keepaspectratio=true]{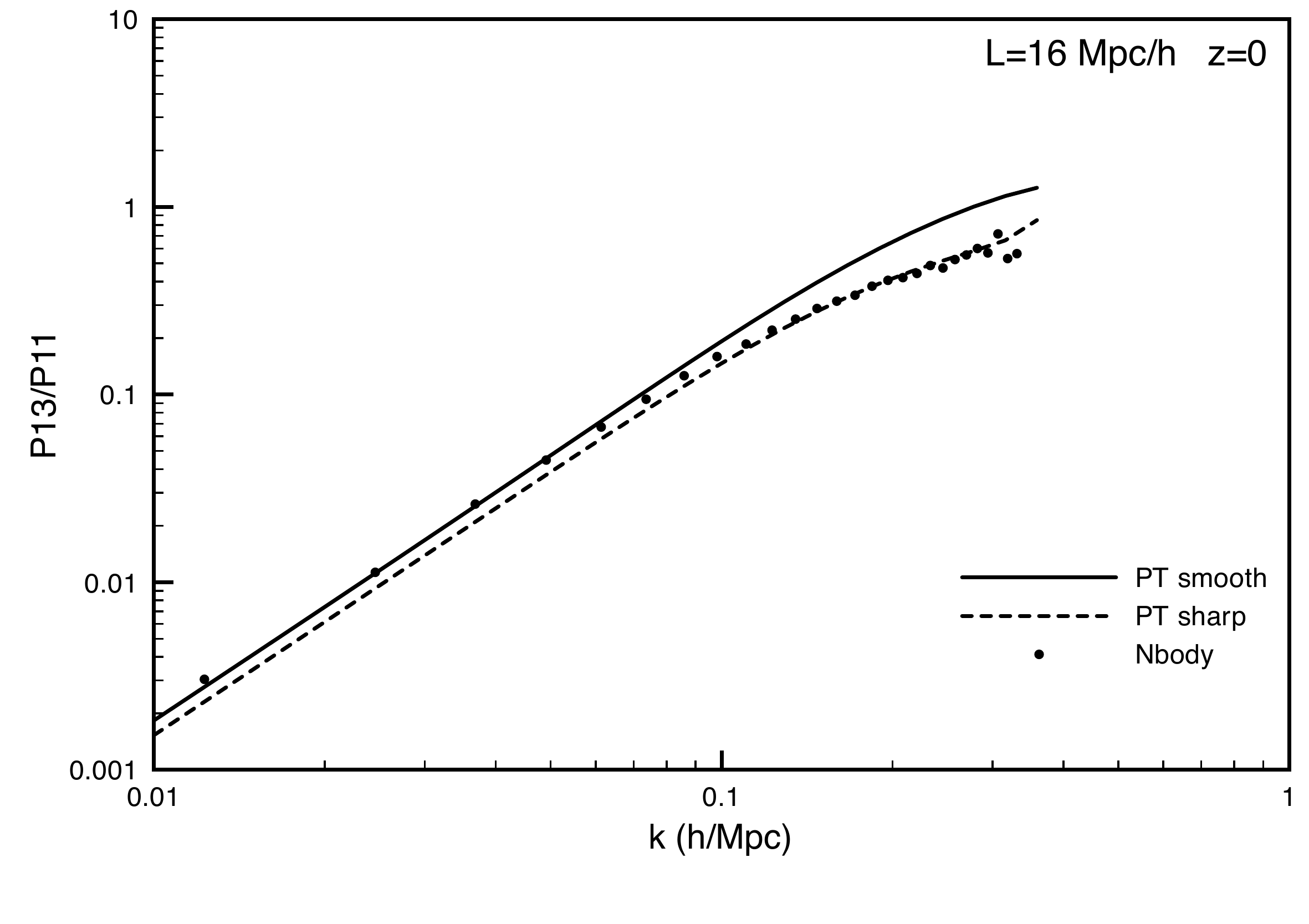}}
\caption{Ratios between $P_{13}$ and $P_{11}$. The continuos (dashed) lines represent the results obtained with the smooth (sharp) cut-off. The dots are obtained from N-body simulations. }
\label{Pratios}
\end{figure}

Finally, in order to estimate the importance of the dynamics at short scales on the coarse-grained quantities, in Figure~\ref{Ph} we plot again $P_{11}$ as obtained in PT (solid line) and from simulations (dots), along with the $\delta P^{in}$-dependent contributions to eq.~(\ref{p1loop}) (dash-dotted line) -- {\it i.e.} the short-wavelength component of the linear PS -- and the difference between the simulations and the $\bar{P}^{in}$-dependent contributions to eq.~(\ref{p1loop}) (open squares). In other words, comparing the squares and the dash-dotted line one can estimate the importance of all the short-distance physics not included in the PT result, namely, nonlinear effects beyond 1-loop order and deviations from the single stream approximation.

\begin{figure}
%\centerline{\includegraphics[width = 8cm,keepaspectratio=true]{Corr_R8_z15.pdf}
%\includegraphics[width = 8cm,keepaspectratio=true]{Corr_R16_z15.pdf}}
\centerline{\includegraphics[width = 8cm,keepaspectratio=true]{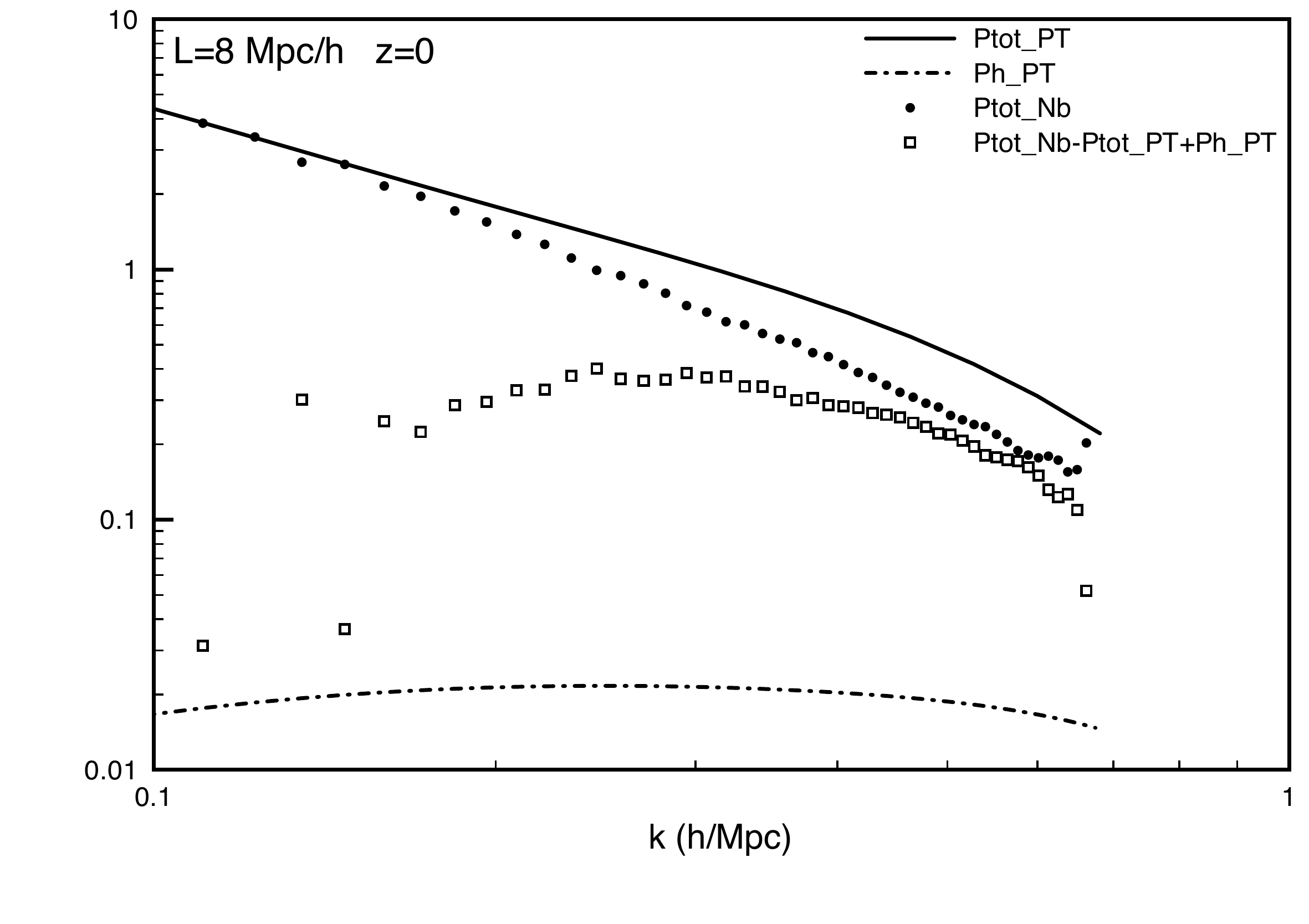}
\includegraphics[width = 8cm,keepaspectratio=true]{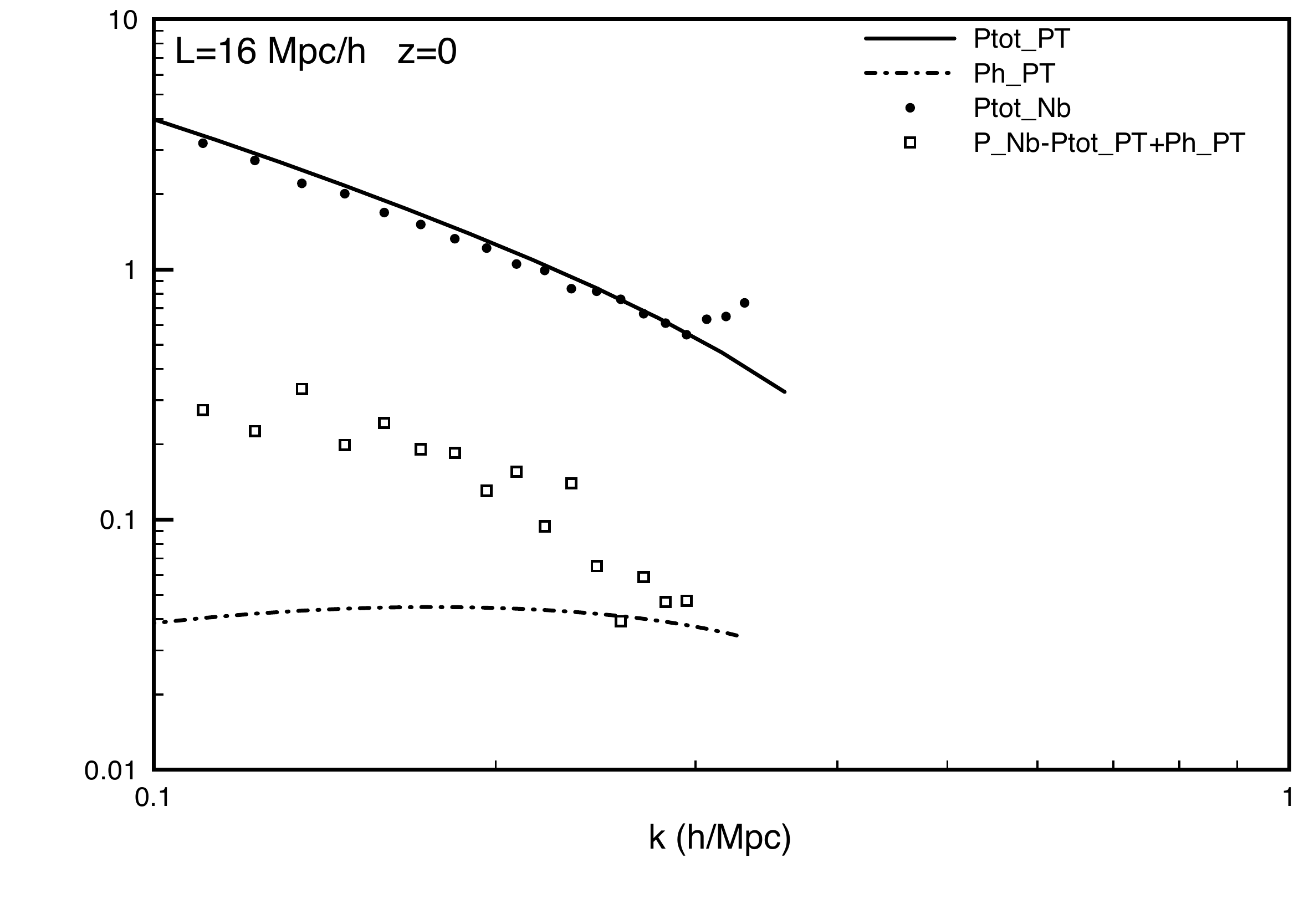}}
\caption{The continuous lines and the dots are the results for $P_{11}$ with the same symbols as in Figure~\ref{Ptutti}. The dash-dotted line and the open squares represent the short-distance contribution to the perturbative result and from simulations, respectively. }
\label{Ph}
\end{figure}

\section{Conclusions}
\label{Conc}

PT and its improvements by means of resummations are a fundamental tool to compare cosmological models to observations and will be even more necessary in the near future. Therefore, any investigation of their shortcomings and possible improvements are welcome. The single stream approximation and some level of coarse-graining are at the basis of these approaches. The impact of these theoretical assumptions on the physical results has been investigated only sporadically \cite{Afshordi:2006ch,Pueblas:2008uv,Valageas:2010rx}. The formulation proposed in this paper provides a solid and clear framework to discuss these approximations.

PT, and its variants, can work only for density fields not too far from linearity; at low redshifts, such smooth fields can be achieved only through coarse-graining up to a scale in the few Mpc range. This inevitably generates a departure from the single stream approximation, which should be taken into account.
Our approach allows a neat separation of modes: by taking an appropriate coarse-graining scale, the long wavelength modes are in the perturbative realm, and their dynamics feels the effect of short distance --non-perturbative-- modes via external sources terms.

In this paper, we have presented a first explicit example of how to deal with these sources, namely, to expand them in PT. This provides a check that the ``traditional'' 1-loop result of PT is recovered (in the sharp cut-off limit) and shows that --even in a fully PT computation -- velocity dispersion emerges macroscopically as a consequence of coarse-graining, and is therefore not exclusively linked to the microscopic phenomenon of shell-crossing. The results obtained in this simple scheme are in (surprising ?) good agreement with simulations.

The next step is to extract and characterize these sources from simulations. It will be relevant to see to what extent the short-distance information encoded in these sources is cosmology-independent. This would open the path to a fast way  to compute the nonlinear PS for different cosmological models: treat the cosmology-dependent long distance modes in PT, and compute the non-perturbative cosmology-independent part once for all in a accurate N-body simulations.

An alternative approach would be to pursue further the analogy with the Wilsonian idea of coarse-graining, and to formulate a Renormalization Group flow by promoting the coarse-graining scale $L$ to the flow parameter: it will be the ideal tool to discuss how short-distance features impact on the long-distance physics, and therefore to extract the --hopefully few-- ``relevant" parameters linked to the short-distance modes.

\section{Acknowledgments} GM is supported by INFN under grant Iniziativa Specifica FA51. GM and MP acknowledge support from grant PRIN-MIUR {\it ``Fisica Astroparticellare"}.

The work of NS was supported by the German Science Foundation (DFG) within the Collaborative Research Center 676 {\it ``Particles, Strings and the Early UniverseÓ}.

MV acknowledges support from the European Commissions FP7 Marie Curie Initial Training Network CosmoComp
(PITN-GA-2009-238356) and by the following grants:
PRIN-INAF 2009, PRIN-MIUR, ASI/AAE, INFN-PD51 and the ERC-Starting
Grant ``cosmoIGM". 

The simulation used was performed at the Darwin
Supercomputer of the University of Cambridge High Performance
Computing Service (http://www.hpc.cam.ac.uk/), provided by Dell Inc.
using Strategic Research Infrastructure Funding from the Higher
Education Funding Council for England.
\section{Appendix A}

We list here the explicit expressions of the vertex functions entering the evolution equation for the four component fields $\bar\vp_a(\bk,\eta)$ defined in Section \ref{compact}
\beqra
&&\gamma_{112}(k,q_1,q_2)=\frac{k^2+q_2^2-q_1^2}{4 q_2^2}\,, \nonumber\\
&&\gamma_{222}(k,q_1,q_2)=\frac{k^2(k^2-q_1^2-q_2^2)}{4 q_1^2 q_2^2}  \,,\nonumber\\
&&\gamma_{213}(k,q_1,q_2)=-\frac{4 q_1^2 q_2^2-(k^2-q_1^2-q_2^2)^2}{16 q_2^4} \,, \nonumber \\
&&\gamma_{214}(k,q_1,q_2)=-\frac{4 q_2^2 k^2-8 q_1^2 q_2^2-4 q_2^4+3(k^2-q_1^2-q_2^2)^2)}{16 q_2^4} \,, \nonumber\\
&&\gamma_{323}(k,q_1,q_2)=-\frac{k^2(k^4+q_1^4+3q_2^4-2k^2(q_1^2+2 q_2^2))}{8 q_1^2 q_2^4}\,, \nonumber\\
&&\gamma_{324}(k,q_1,q_2)=\frac{1}{2}\frac{k^2}{q_2^2}\left(\frac{3(k^2-q_1^2-q_2^2)^2}{4 q_1^2 q_2^2 }-1\right)\,, \nonumber\\
&&\gamma_{423}(k,q_1,q_2)=-\frac{(3k^2+q_1^2-3q_2^2)(k^4+(q_1^2-q_2^2)^2-2k^2(q_1^2+q_2^2))}{32 q_1^2 q_2^4 }\,, \nonumber\\
&&\gamma_{424}(k,q_1,q_2)=\nonumber\\
&&\frac{9k^6+(q_1^2-q_2^2)^2(3q_1^2-q_2^2)-k^4(15 q_1^2+11 q_2^2)+k^2(3q_1^4+2q_1^2q_2^2+3q_2^4)}{32 q_1^2 q_2^4 }\,, \nonumber\\
\eeqra
with, finally,  $\gamma_{abc}(k,q_1,q_2)=\gamma_{acb}(k,q_2,q_1)$.

\section{Appendix B}

Using eqs. (\ref{neweul}) and (\ref{newsig}) one finds that the three non vanishing components of the source vector $h_a(\bk,\eta)$ are the following
\beqra
&& h_2(\bk,\eta) = -\frac{3}{2}\frac{\Omega_m}{f^2}  e^\eta\, \tilde{W}(\bk L) \int d^3q \,\frac{\bk\cdot\bq}{q^2}\,\delta\vp_1(\bq,\eta)\,\delta\vp_1(\bk-\bq,\eta) \,,
\label{h2} \\
&& h_{3}(\bk,\eta) = 3\frac{\Omega_m}{f^2} e^\eta\, \tilde{W}(\bk L)  \int d^3q \; k^2 \frac{\bq \cdot (\bk-\bq)}{q^2 |\bk-\bq|^2}  \delta\vp_1(\bq,\eta)\,\delta\vp_2(\bk-\bq,\eta)\,, \nonumber \\
&& \label{h3} \\
&& h_{4}(\bk,\eta) = 3\frac{\Omega_m}{f^2} e^\eta\, \tilde{W}(\bk L) \int d^3q \;  \frac{\bk \cdot \bq (k^2-\bk \cdot\bq)}{q^2 |\bk-\bq|^2}  \delta\vp_1(\bq,\eta)\,\delta\vp_2(\bk-\bq,\eta)\,. \nonumber \\
&&
 \label{h34}
\eeqra

\section*{References}
\bibliographystyle{JHEP}
\bibliography{mybib.bib}

\providecommand{\href}[2]{#2}\begingroup\raggedright\begin{thebibliography}{10}

\bibitem{Blake:2011wn}
C.~Blake, T.~Davis, G.~Poole, D.~Parkinson, S.~Brough {\em et.~al.}, {\it {The
  WiggleZ Dark Energy Survey: testing the cosmological model with baryon
  acoustic oscillations at z=0.6}},  \href{http://arXiv.org/abs/1105.2862}{{\tt
  1105.2862}}.

\bibitem{Blake:2011rj}
C.~Blake, S.~Brough, M.~Colless, C.~Contreras, W.~Couch {\em et.~al.}, {\it
  {The WiggleZ Dark Energy Survey: the growth rate of cosmic structure since
  redshift z=0.9}},  \href{http://arXiv.org/abs/1104.2948}{{\tt 1104.2948}}.

\bibitem{Eisenstein:2011sa}
{\bf SDSS Collaboration} Collaboration, D.~J. Eisenstein {\em et.~al.}, {\it
  {SDSS-III: Massive Spectroscopic Surveys of the Distant Universe, the Milky
  Way Galaxy, and Extra-Solar Planetary Systems}},  {\em Astron.J.} (2011)
  [\href{http://arXiv.org/abs/1101.1529}{{\tt 1101.1529}}].

\bibitem{Bassett:2005kn}
{\bf WFMOS Collaboration} Collaboration, B.~A. Bassett, R.~C. Nichol and D.~J.
  Eisenstein, {\it {WFMOS: Sounding the dark cosmos}},  {\em Astron.Geophys.}
  {\bf 46} (2005) 526--529 [\href{http://arXiv.org/abs/astro-ph/0510272}{{\tt
  astro-ph/0510272}}].

\bibitem{2008ASPC..399..115H}
G.~J. {Hill}, K.~{Gebhardt}, E.~{Komatsu}, N.~{Drory}, P.~J. {MacQueen},
  J.~{Adams}, G.~A. {Blanc}, R.~{Koehler}, M.~{Rafal}, M.~M. {Roth}, A.~{Kelz},
  C.~{Gronwall}, R.~{Ciardullo} and D.~P. {Schneider}, {\it {The Hobby-Eberly
  Telescope Dark Energy Experiment (HETDEX): Description and Early Pilot Survey
  Results}},  in {\em Panoramic Views of Galaxy Formation and Evolution}
  ({T.~Kodama, T.~Yamada, \& K.~Aoki}, ed.), vol.~399 of {\em Astronomical
  Society of the Pacific Conference Series}, pp.~115--+, 2008.
\newblock \href{http://arXiv.org/abs/0806.0183}{{\tt 0806.0183}}.

\bibitem{2009arXiv0912.0914L}
R.~{Laureijs}, {\it {Euclid Assessment Study Report for the ESA Cosmic
  Visions}},  \href{http://arXiv.org/abs/0912.0914}{{\tt 0912.0914}}.

\bibitem{2009arXiv0901.0721A}
A.~{Albrecht}, L.~{Amendola}, G.~{Bernstein}, D.~{Clowe}, D.~{Eisenstein},
  L.~{Guzzo}, C.~{Hirata}, D.~{Huterer}, R.~{Kirshner}, E.~{Kolb} and
  R.~{Nichol}, {\it {Findings of the Joint Dark Energy Mission Figure of Merit
  Science Working Group}},  \href{http://arXiv.org/abs/0901.0721}{{\tt
  0901.0721}}.

\bibitem{Percival:2009xn}
{\bf SDSS Collaboration} Collaboration, B.~A. Reid {\em et.~al.}, {\it {Baryon
  Acoustic Oscillations in the Sloan Digital Sky Survey Data Release 7 Galaxy
  Sample}},  {\em Mon.Not.Roy.Astron.Soc.} {\bf 401} (2010) 2148--2168
  [\href{http://arXiv.org/abs/0907.1660}{{\tt 0907.1660}}].

\bibitem{Schlegel:2011wb}
{\bf BigBOSS Collaboration} Collaboration, D.~Schlegel {\em et.~al.}, {\it {The
  BigBOSS Experiment}},  \href{http://arXiv.org/abs/1106.1706}{{\tt
  1106.1706}}.

\bibitem{Eis05}
{\bf SDSS} Collaboration, D.~J. Eisenstein {\em et.~al.}, {\it {Detection of
  the Baryon Acoustic Peak in the Large-Scale Correlation Function of SDSS
  Luminous Red Galaxies}},  {\em Astrophys. J.} {\bf 633} (2005) 560--574
  [\href{http://arXiv.org/abs/astro-ph/0501171}{{\tt astro-ph/0501171}}].
%%CITATION = ASTRO-PH/0501171;%%

\bibitem{2007ApJ...657...51P}
W.~J. {Percival}, R.~C. {Nichol}, D.~J. {Eisenstein}, D.~H. {Weinberg},
  M.~{Fukugita}, A.~C. {Pope}, D.~P. {Schneider}, A.~S. {Szalay}, M.~S.
  {Vogeley}, I.~{Zehavi}, N.~A. {Bahcall}, J.~{Brinkmann}, A.~J. {Connolly},
  J.~{Loveday} and A.~{Meiksin}, {\it {Measuring the Matter Density Using
  Baryon Oscillations in the SDSS}},  {\em Astrophys. J.} {\bf 657} (Mar.,
  2007) 51--55 [\href{http://arXiv.org/abs/arXiv:astro-ph/0608635}{{\tt
  arXiv:astro-ph/0608635}}].

\bibitem{2007MNRAS.381.1053P}
W.~J. {Percival}, S.~{Cole}, D.~J. {Eisenstein}, R.~C. {Nichol}, J.~A.
  {Peacock}, A.~C. {Pope} and A.~S. {Szalay}, {\it {Measuring the Baryon
  Acoustic Oscillation scale using the Sloan Digital Sky Survey and 2dF Galaxy
  Redshift Survey}},  {\em Mon.Not.Roy.Astron.Soc.} {\bf 381} (Nov., 2007)
  1053--1066 [\href{http://arXiv.org/abs/0705.3323}{{\tt 0705.3323}}].

\bibitem{2009MNRAS.399.1663G}
E.~{Gazta{\~n}aga}, A.~{Cabr{\'e}} and L.~{Hui}, {\it {Clustering of luminous
  red galaxies - IV. Baryon acoustic peak in the line-of-sight direction and a
  direct measurement of H(z)}},  {\em Mon.Not.Roy.Astron.Soc.} {\bf 399} (Nov.,
  2009) 1663--1680 [\href{http://arXiv.org/abs/0807.3551}{{\tt 0807.3551}}].

\bibitem{Abazajian:2011dt}
K.~Abazajian, E.~Calabrese, A.~Cooray, F.~De~Bernardis, S.~Dodelson {\em
  et.~al.}, {\it {Cosmological and Astrophysical Neutrino Mass Measurements}},
  \href{http://arXiv.org/abs/1103.5083}{{\tt 1103.5083}}.

\bibitem{Heitmann:2008eq}
K.~Heitmann, M.~White, C.~Wagner, S.~Habib and D.~Higdon, {\it {The Coyote
  Universe I: Precision Determination of the Nonlinear Matter Power Spectrum}},
   {\em Astrophys.J.} {\bf 715} (2010) 104--121
  [\href{http://arXiv.org/abs/0812.1052}{{\tt 0812.1052}}].

\bibitem{Heitmann:2009cu}
K.~Heitmann, D.~Higdon, M.~White, S.~Habib, B.~J. Williams {\em et.~al.}, {\it
  {The Coyote Universe II: Cosmological Models and Precision Emulation of the
  Nonlinear Matter Power Spectrum}},  {\em Astrophys.J.} {\bf 705} (2009)
  156--174 [\href{http://arXiv.org/abs/0902.0429}{{\tt 0902.0429}}].

\bibitem{Lawrence:2009uk}
E.~Lawrence, K.~Heitmann, M.~White, D.~Higdon, C.~Wagner {\em et.~al.}, {\it
  {The Coyote Universe III: Simulation Suite and Precision Emulator for the
  Nonlinear Matter Power Spectrum}},  {\em Astrophys.J.} {\bf 713} (2010)
  1322--1331 [\href{http://arXiv.org/abs/0912.4490}{{\tt 0912.4490}}].

\bibitem{Viel:2011bk}
M.~Viel, K.~Markovic, M.~Baldi and J.~Weller, {\it {The Non-Linear Matter Power
  Spectrum in Warm Dark Matter Cosmologies}},
  \href{http://arXiv.org/abs/1107.4094}{{\tt 1107.4094}}.
%%CITATION = 1107.4094;%%

\bibitem{PT}
F.~Bernardeau, S.~Colombi, E.~Gaztanaga and R.~Scoccimarro, {\it {Large-scale
  structure of the universe and cosmological perturbation theory}},  {\em Phys.
  Rept.} {\bf 367} (2002) 1--248
  [\href{http://arXiv.org/abs/astro-ph/0112551}{{\tt astro-ph/0112551}}].
%%CITATION = ASTRO-PH/0112551;%%

\bibitem{JK06}
D.~Jeong and E.~Komatsu, {\it {Perturbation Theory Reloaded: Analytical
  Calculation of Non-linearity in Baryonic Oscillations in the Real Space
  Matter Power Spectrum}},  {\em Astrophys. J.} {\bf 651} (2006) 619
  [\href{http://arXiv.org/abs/astro-ph/0604075}{{\tt astro-ph/0604075}}].
%%CITATION = ASTRO-PH/0604075;%%

\bibitem{STT08}
S.~Saito, M.~Takada and A.~Taruya, {\it {Impact of massive neutrinos on
  nonlinear matter power spectrum}},  {\em Phys. Rev. Lett.} {\bf 100} (2008)
  191301 [\href{http://arXiv.org/abs/0801.0607}{{\tt 0801.0607}}].
%%CITATION = 0801.0607;%%

\bibitem{Wong08}
Y.~Y.~Y. Wong, {\it {Higher order corrections to the large scale matter power
  spectrum in the presence of massive neutrinos}},
  \href{http://arXiv.org/abs/0809.0693}{{\tt 0809.0693}}.
%%CITATION = 0809.0693;%%

\bibitem{SaitoII}
S.~Saito, M.~Takada and A.~Taruya, {\it {Nonlinear power spectrum in the
  presence of massive neutrinos: perturbation theory approach, galaxy bias and
  parameter forecasts}},  {\em Phys. Rev.} {\bf D80} (2009) 083528
  [\href{http://arXiv.org/abs/0907.2922}{{\tt 0907.2922}}].
%%CITATION = 0907.2922;%%

\bibitem{Sefusatti:2009qh}
E.~Sefusatti, {\it {1-loop Perturbative Corrections to the Matter and Galaxy
  Bispectrum with non-Gaussian Initial Conditions}},  {\em Phys.Rev.} {\bf D80}
  (2009) 123002 [\href{http://arXiv.org/abs/0905.0717}{{\tt 0905.0717}}].

\bibitem{RPTa}
M.~Crocce and R.~Scoccimarro, {\it {Renormalized Cosmological Perturbation
  Theory}},  {\em Phys. Rev.} {\bf D73} (2006) 063519
  [\href{http://arXiv.org/abs/astro-ph/0509418}{{\tt astro-ph/0509418}}].
%%CITATION = ASTRO-PH/0509418;%%

\bibitem{RPTb}
M.~Crocce and R.~Scoccimarro, {\it {Memory of Initial Conditions in
  Gravitational Clustering}},  {\em Phys. Rev.} {\bf D73} (2006) 063520
  [\href{http://arXiv.org/abs/astro-ph/0509419}{{\tt astro-ph/0509419}}].
%%CITATION = ASTRO-PH/0509419;%%

\bibitem{Valageas03}
P.~Valageas, {\it {A new approach to gravitational clustering: a path- integral
  formalism and large-N expansions}},  {\em Astron. Astrophys.} {\bf 421}
  (2004) 23--40 [\href{http://arXiv.org/abs/astro-ph/0307008}{{\tt
  astro-ph/0307008}}].
%%CITATION = ASTRO-PH/0307008;%%

\bibitem{McD06}
P.~McDonald, {\it {Dark matter clustering: a simple renormalization group
  approach}},  {\em Phys. Rev.} {\bf D75} (2007) 043514
  [\href{http://arXiv.org/abs/astro-ph/0606028}{{\tt astro-ph/0606028}}].
%%CITATION = ASTRO-PH/0606028;%%

\bibitem{MP07b}
S.~Matarrese and M.~Pietroni, {\it {Resumming Cosmic Perturbations}},  {\em
  JCAP} {\bf 0706} (2007) 026
  [\href{http://arXiv.org/abs/astro-ph/0703563}{{\tt astro-ph/0703563}}].
%%CITATION = ASTRO-PH/0703563;%%

\bibitem{Matsubara07}
T.~Matsubara, {\it {Resumming Cosmological Perturbations via the Lagrangian
  Picture: One-loop Results in Real Space and in Redshift Space}},  {\em Phys.
  Rev.} {\bf D77} (2008) 063530 [\href{http://arXiv.org/abs/0711.2521}{{\tt
  0711.2521}}].
%%CITATION = 0711.2521;%%

\bibitem{Taruya2007}
A.~{Taruya} and T.~{Hiramatsu}, {\it {A Closure Theory for Nonlinear Evolution
  of Cosmological Power Spectra}},  {\em Astrophys.J.} {\bf 674} (Feb., 2008)
  617--635 [\href{http://arXiv.org/abs/0708.1367}{{\tt 0708.1367}}].

\bibitem{Pietroni08}
M.~Pietroni, {\it {Flowing with Time: a New Approach to Nonlinear Cosmological
  Perturbations}},  {\em JCAP} {\bf 0810} (2008) 036
  [\href{http://arXiv.org/abs/0806.0971}{{\tt 0806.0971}}].
%%CITATION = 0806.0971;%%

\bibitem{TaruyaIII}
T.~Hiramatsu and A.~Taruya, {\it {Chasing the non-linear evolution of matter
  power spectrum with numerical resummation method: solution of closure
  equations}},  {\em Phys. Rev.} {\bf D79} (2009) 103526
  [\href{http://arXiv.org/abs/0902.3772}{{\tt 0902.3772}}].
%%CITATION = 0902.3772;%%

\bibitem{MP07a}
S.~Matarrese and M.~Pietroni, {\it {Baryonic acoustic oscillations via the
  renormalization group}},  {\em Mod. Phys. Lett.} {\bf A23} (2008) 25--32
  [\href{http://arXiv.org/abs/astro-ph/0702653}{{\tt astro-ph/0702653}}].
%%CITATION = ASTRO-PH/0702653;%%

\bibitem{RPTBAO}
M.~Crocce and R.~Scoccimarro, {\it {Nonlinear Evolution of Baryon Acoustic
  Oscillations}},  {\em Phys. Rev.} {\bf D77} (2008) 023533
  [\href{http://arXiv.org/abs/0704.2783}{{\tt 0704.2783}}].
%%CITATION = 0704.2783;%%

\bibitem{Taruya:2009ir}
A.~Taruya, T.~Nishimichi, S.~Saito and T.~Hiramatsu, {\it {Non-linear Evolution
  of Baryon Acoustic Oscillations from Improved Perturbation Theory in Real and
  Redshift Spaces}},  \href{http://arXiv.org/abs/0906.0507}{{\tt 0906.0507}}.
%%CITATION = 0906.0507;%%

\bibitem{Taruya:2010mx}
A.~Taruya, T.~Nishimichi and S.~Saito, {\it {Baryon Acoustic Oscillations in
  2D: Modeling Redshift- space Power Spectrum from Perturbation Theory}},  {\em
  Phys. Rev.} {\bf D82} (2010) 063522
  [\href{http://arXiv.org/abs/1006.0699}{{\tt 1006.0699}}].
%%CITATION = 1006.0699;%%

\bibitem{LMPR09}
J.~Lesgourgues, S.~Matarrese, M.~Pietroni and A.~Riotto, {\it {Non-linear Power
  Spectrum including Massive Neutrinos: the Time-RG Flow Approach}},  {\em
  JCAP} {\bf 0906} (2009) 017 [\href{http://arXiv.org/abs/0901.4550}{{\tt
  0901.4550}}].
%%CITATION = 0901.4550;%%

\bibitem{Bartolo:2009rb}
N.~Bartolo, J.~P.~B. Almeida, S.~Matarrese, M.~Pietroni and A.~Riotto, {\it
  {Signatures of Primordial non-Gaussianities in the Matter Power-Spectrum and
  Bispectrum: the Time-RG Approach}},  {\em JCAP} {\bf 1003} (2010) 011
  [\href{http://arXiv.org/abs/0912.4276}{{\tt 0912.4276}}].
%%CITATION = 0912.4276;%%

\bibitem{SefuDami}
G.~{D'Amico} and E.~{Sefusatti}, {\it {The nonlinear power spectrum in
  clustering quintessence cosmologies}},
  \href{http://arXiv.org/abs/1106.0314}{{\tt 1106.0314}}.

\bibitem{Anselmi:2011ef}
S.~Anselmi, G.~Ballesteros and M.~Pietroni, {\it {Non-linear dark energy
  clustering}},  \href{http://arXiv.org/abs/1106.0834}{{\tt 1106.0834}}.
%%CITATION = 1106.0834;%%

\bibitem{Carlson:2009it}
J.~Carlson, M.~White and N.~Padmanabhan, {\it {A critical look at cosmological
  perturbation theory techniques}},  {\em Phys. Rev.} {\bf D80} (2009) 043531
  [\href{http://arXiv.org/abs/0905.0479}{{\tt 0905.0479}}].
%%CITATION = 0905.0479;%%

\bibitem{Bernardeau:2008fa}
F.~Bernardeau, M.~Crocce and R.~Scoccimarro, {\it {Multi-Point Propagators in
  Cosmological Gravitational Instability}},  {\em Phys.Rev.} {\bf D78} (2008)
  103521 [\href{http://arXiv.org/abs/0806.2334}{{\tt 0806.2334}}].

\bibitem{Anselmi:2010fs}
S.~Anselmi, S.~Matarrese and M.~Pietroni, {\it {Next-to-leading resummations in
  cosmological perturbation theory}},  {\em JCAP} {\bf 1106} (2011) 015
  [\href{http://arXiv.org/abs/1011.4477}{{\tt 1011.4477}}].

\bibitem{Pueblas:2008uv}
S.~Pueblas and R.~Scoccimarro, {\it {Generation of Vorticity and Velocity
  Dispersion by Orbit Crossing}},  {\em Phys.Rev.} {\bf D80} (2009) 043504
  [\href{http://arXiv.org/abs/0809.4606}{{\tt 0809.4606}}].

\bibitem{Valageas:2010rx}
P.~Valageas, {\it {Impact of shell crossing and scope of perturbative
  approaches in real and redshift space}},  {\em Astron. Astrophys.} {\bf 526}
  (2011) A67 [\href{http://arXiv.org/abs/1009.0106}{{\tt 1009.0106}}].
%%CITATION = 1009.0106;%%

\bibitem{Baumann:2010tm}
D.~Baumann, A.~Nicolis, L.~Senatore and M.~Zaldarriaga, {\it {Cosmological
  Non-Linearities as an Effective Fluid}},
  \href{http://arXiv.org/abs/1004.2488}{{\tt 1004.2488}}.
%%CITATION = 1004.2488;%%

\bibitem{Buchert:2005xj}
T.~Buchert and A.~Dominguez, {\it {Adhesive Gravitational Clustering}},  {\em
  Astron. Astrophys.} {\bf 438} (2005) 443--460
  [\href{http://arXiv.org/abs/astro-ph/0502318}{{\tt astro-ph/0502318}}].
%%CITATION = ASTRO-PH/0502318;%%

\bibitem{McDonald:2009hs}
P.~McDonald, {\it {How to generate a significant effective temperature for cold
  dark matter, from first principles}},  {\em JCAP} {\bf 1104} (2011) 032
  [\href{http://arXiv.org/abs/0910.1002}{{\tt 0910.1002}}].

\bibitem{volker05}
V.~Springel, {\it {The Cosmological simulation code GADGET-2}},  {\em
  Mon.Not.Roy.Astron.Soc.} {\bf 364} (2005) 1105--1134
  [\href{http://arXiv.org/abs/astro-ph/0505010}{{\tt astro-ph/0505010}}].

\bibitem{eisensteinhu}
D.~J. Eisenstein and W.~Hu, {\it {Power spectra for cold dark matter and its
  variants}},  {\em Astrophys.J.} {\bf 511} (1997) 5
  [\href{http://arXiv.org/abs/astro-ph/9710252}{{\tt astro-ph/9710252}}].

\bibitem{Afshordi:2006ch}
N.~Afshordi, {\it {How well can (renormalized) perturbation theory predict dark
  matter clustering properties?}},  {\em Phys.Rev.} {\bf D75} (2007) 021302
  [\href{http://arXiv.org/abs/astro-ph/0610336}{{\tt astro-ph/0610336}}].

\end{thebibliography}\endgroup
%\bibliography{/Users/pietroni/Bibliografia/mybib.bib}
%\bibliography{/Utenti/gianpiero/Scrivania/SIDM/new/mybib.bib}
\end{document}